\pdfoutput=1

\documentclass[12pt,a4paper]{article}

\usepackage{ifthen} 
\newboolean{pdflatex}
\setboolean{pdflatex}{true} 

\newboolean{articletitles}
\setboolean{articletitles}{true} 

\newboolean{uprightparticles}
\setboolean{uprightparticles}{false} 

\def\paperauthors{LHCb collaboration} 
\def\paperasciititle{Measurement of psi(2S) meson production in pp collisions at 7 TeV} 
\def\papertitle{Measurement of $\psitwos$ meson production in $pp$ collisions at $\sqrt{s}=7$~TeV} 
\def\paperkeywords{{High Energy Physics}, {LHCb}} 
\def\papercopyright{\the\year\ CERN for the benefit of the LHCb collaboration} 
\def\paperlicence{CC-BY-4.0 licence}
\def\paperlicenceurl{https://creativecommons.org/licenses/by/4.0/}

\usepackage[top=1in, bottom=1.25in, left=1in, right=1in]{geometry}

\usepackage{microtype}
\usepackage{lineno}  
\usepackage{xspace} 
\usepackage{caption} 

\usepackage{graphicx}  
\usepackage{color}
\usepackage{colortbl}
\graphicspath{{./figs/}} 
\DeclareGraphicsExtensions{.pdf,.PDF,png,.PNG}

\usepackage{amsmath} 
\usepackage{amssymb}
\usepackage{amsfonts}
\usepackage{upgreek} 

\newcommand*\patchAmsMathEnvironmentForLineno[1]{%
\expandafter\let\csname old#1\expandafter\endcsname\csname #1\endcsname
\expandafter\let\csname oldend#1\expandafter\endcsname\csname
end#1\endcsname
 \renewenvironment{#1}%
   {\linenomath\csname old#1\endcsname}%
   {\csname oldend#1\endcsname\endlinenomath}%
}
\newcommand*\patchBothAmsMathEnvironmentsForLineno[1]{%
  \patchAmsMathEnvironmentForLineno{#1}%
  \patchAmsMathEnvironmentForLineno{#1*}%
}
\AtBeginDocument{%
\patchBothAmsMathEnvironmentsForLineno{equation}%
\patchBothAmsMathEnvironmentsForLineno{align}%
\patchBothAmsMathEnvironmentsForLineno{flalign}%
\patchBothAmsMathEnvironmentsForLineno{alignat}%
\patchBothAmsMathEnvironmentsForLineno{gather}%
\patchBothAmsMathEnvironmentsForLineno{multline}%
\patchBothAmsMathEnvironmentsForLineno{eqnarray}%
}

\usepackage{hyperxmp}

\usepackage[pdftex,
            pdfauthor={\paperauthors},
            pdftitle={\paperasciititle},
            pdfkeywords={\paperkeywords},
            pdfcopyright={Copyright (C) \papercopyright},
            pdflicenseurl={\paperlicenceurl}]{hyperref}

\usepackage[colorinlistoftodos,textsize=scriptsize]{todonotes}

\usepackage[all]{hypcap} 

\usepackage{xspace} 
\usepackage{upgreek}

\def\lhcb   {\mbox{LHCb}\xspace}

\def\MagUp {\mbox{\em Mag\kern -0.05em Up}\xspace}

\ifthenelse{\boolean{uprightparticles}}%
{

 \def\Pmu         {\ensuremath{\upmu}\xspace}

 \def\Ppi         {\ensuremath{\uppi}\xspace}

 \def\Ppsi        {\ensuremath{\uppsi}\xspace}

 \def\PDelta      {\ensuremath{\Delta}\xspace}                 
 \def\PXi         {\ensuremath{\Xi}\xspace}                 
 \def\PLambda     {\ensuremath{\Lambda}\xspace}                 
 \def\PSigma      {\ensuremath{\Sigma}\xspace}                 
 \def\POmega      {\ensuremath{\Omega}\xspace}                 
 \def\PUpsilon    {\ensuremath{\Upsilon}\xspace}

 \def\PB      {\ensuremath{\mathrm{B}}\xspace}                 
                  
 \def\PD      {\ensuremath{\mathrm{D}}\xspace}

 \def\PK      {\ensuremath{\mathrm{K}}\xspace}

 \def\Pb      {\ensuremath{\mathrm{b}}\xspace}

 \def\Pi      {\ensuremath{\mathrm{i}}\xspace}

 \def\Ps      {\ensuremath{\mathrm{s}}\xspace}

 \def\thebaroffset{0.0em}
}
{

 \def\Pmu         {\ensuremath{\mu}\xspace}

 \def\Ppi         {\ensuremath{\pi}\xspace}

 \def\Ppsi        {\ensuremath{\psi}\xspace}                 
                  
 \mathchardef\PDelta="7101
 \mathchardef\PXi="7104
 \mathchardef\PLambda="7103
 \mathchardef\PSigma="7106
 \mathchardef\POmega="710A
 \mathchardef\PUpsilon="7107
                  
 \def\PB      {\ensuremath{B}\xspace}                 
                  
 \def\PD      {\ensuremath{D}\xspace}

 \def\PK      {\ensuremath{K}\xspace}

 \def\Pb      {\ensuremath{b}\xspace}

 \def\Pi      {\ensuremath{i}\xspace}

 \def\Ps      {\ensuremath{s}\xspace}

 \def\thebaroffset{0.18em}
}
\newcommand{\offsetoverline}[2][\thebaroffset]{\kern #1\overline{\kern -#1 #2}}%

\makeatletter
\ifcase \@ptsize \relax
  \newcommand{\miniscule}{\@setfontsize\miniscule{4}{5}}
\or
  \newcommand{\miniscule}{\@setfontsize\miniscule{5}{6}}
\or
  \newcommand{\miniscule}{\@setfontsize\miniscule{5}{6}}
\fi
\makeatother

\DeclareRobustCommand{\optbar}[1]{\shortstack{{\miniscule (\rule[.5ex]{1.25em}{.18mm})}
  \\ [-.7ex] $#1$}}

\def\mumu       {{\ensuremath{\Pmu^+\Pmu^-}}\xspace}

\def\squark    {{\ensuremath{\Ps}}\xspace}

\def\bquark    {{\ensuremath{\Pb}}\xspace}

\def\pion   {{\ensuremath{\Ppi}}\xspace}

\def\pip    {{\ensuremath{\pion^+}}\xspace}
\def\pim    {{\ensuremath{\pion^-}}\xspace}

\def\KorKbar {\kern \thebaroffset\optbar{\kern -\thebaroffset \PK}{}\xspace}

\def\DorDbar {\kern \thebaroffset\optbar{\kern -\thebaroffset \PD}\xspace}

\def\B       {{\ensuremath{\PB}}\xspace}

\def\BorBbar {\kern \thebaroffset\optbar{\kern -\thebaroffset \PB}\xspace}

\def\Bd      {{\ensuremath{\B^0}}\xspace}

\def\BdorBdbar {\kern \thebaroffset\optbar{\kern -\thebaroffset \Bd}\xspace}

\def\Bs      {{\ensuremath{\B^0_\squark}}\xspace}

\def\BsorBsbar {\kern \thebaroffset\optbar{\kern -\thebaroffset \Bs}\xspace}

\def\jpsi     {{\ensuremath{{\PJ\mskip -3mu/\mskip -2mu\Ppsi\mskip 2mu}}}\xspace}
\def\psitwos  {{\ensuremath{\Ppsi{(2S)}}}\xspace}

\def\Y#1S{\ensuremath{\PUpsilon{(#1S)}}\xspace}

\def\Lz          {{\ensuremath{\PLambda}}\xspace}

\def\LorLbar     {\kern \thebaroffset\optbar{\kern -\thebaroffset \PLambda}\xspace}

\def\Lb           {{\ensuremath{\Lz^0_\bquark}}\xspace}

\def\BF         {{\ensuremath{\mathcal{B}}}\xspace}
\def\BR         {\BF}

\def\to                 {\ensuremath{\rightarrow}\xspace}

\def\CP                {{\ensuremath{C\!P}}\xspace}

\def\AT#1     {\ensuremath{A_{\mathrm{T}}^{#1}}\xspace}           

\def\C#1      {\ensuremath{\mathcal{C}_{#1}}\xspace}                       
\def\Cp#1     {\ensuremath{\mathcal{C}_{#1}^{'}}\xspace}                    
\def\Ceff#1   {\ensuremath{\mathcal{C}_{#1}^{\mathrm{(eff)}}}\xspace}        
\def\Cpeff#1  {\ensuremath{\mathcal{C}_{#1}^{'\mathrm{(eff)}}}\xspace}       
\def\Ope#1    {\ensuremath{\mathcal{O}_{#1}}\xspace}                       
\def\Opep#1   {\ensuremath{\mathcal{O}_{#1}^{'}}\xspace}                    

\newcommand{\nospaceunit}[1]{\ensuremath{\text{#1}}}       
\newcommand{\aunit}[1]{\ensuremath{\text{\,#1}}}       

\newcommand{\tev}{\aunit{Te\kern -0.1em V}\xspace}
\newcommand{\gev}{\aunit{Ge\kern -0.1em V}\xspace}
\newcommand{\mev}{\aunit{Me\kern -0.1em V}\xspace}
\newcommand{\kev}{\aunit{ke\kern -0.1em V}\xspace}
\newcommand{\ev}{\aunit{e\kern -0.1em V}\xspace}
\newcommand{\mevc}{\ensuremath{\aunit{Me\kern -0.1em V\!/}c}\xspace}
\newcommand{\gevc}{\ensuremath{\aunit{Ge\kern -0.1em V\!/}c}\xspace}
\newcommand{\mevcc}{\ensuremath{\aunit{Me\kern -0.1em V\!/}c^2}\xspace}
\newcommand{\gevcc}{\ensuremath{\aunit{Ge\kern -0.1em V\!/}c^2}\xspace}

\def\mm   {\aunit{mm}\xspace}

\def\mub{\ensuremath{\,\upmu\nospaceunit{b}}\xspace}
\def\nb {\aunit{nb}\xspace}

\def\pb {\aunit{pb}\xspace}
\def\invpb {\ensuremath{\pb^{-1}}\xspace}

\def\ps   {\ensuremath{\aunit{ps}}\xspace}

\def\mhz  {\ensuremath{\aunit{MHz}}\xspace}

\def\deriv {\ensuremath{\mathrm{d}}}

\def\gsim{{~\raise.15em\hbox{$>$}\kern-.85em
          \lower.35em\hbox{$\sim$}~}\xspace}
\def\lsim{{~\raise.15em\hbox{$<$}\kern-.85em
          \lower.35em\hbox{$\sim$}~}\xspace}

\def\sqs   {\ensuremath{\protect\sqrt{s}}\xspace}

\def\pt         {\ensuremath{p_{\mathrm{T}}}\xspace}

\newcommand{\lum} {\ensuremath{\mathcal{L}}\xspace}

\def\evtgen     {\mbox{\textsc{EvtGen}}\xspace}

\def\gauss      {\mbox{\textsc{Gauss}}\xspace}
\def\geant      {\mbox{\textsc{Geant4}}\xspace}

\def\photos     {\mbox{\textsc{Photos}}\xspace}

\def\pythia     {\mbox{\textsc{Pythia}}\xspace}

\def\tell1  {TELL1\xspace}
\def\ukl1   {UKL1\xspace}

\newcommand{\etal}{\mbox{\itshape et al.}\xspace}

\usepackage{cite} 
\usepackage{mciteplus}

\usepackage{longtable} 

\begin{document}

\renewcommand{\thefootnote}{\fnsymbol{footnote}}
\setcounter{footnote}{1}


\begin{titlepage}
\pagenumbering{roman}

\vspace*{-1.5cm}
\centerline{\large EUROPEAN ORGANIZATION FOR NUCLEAR RESEARCH (CERN)}
\vspace*{1.5cm}
\noindent
\begin{tabular*}{\linewidth}{lc@{\extracolsep{\fill}}r@{\extracolsep{0pt}}}
\ifthenelse{\boolean{pdflatex}}
{\vspace*{-1.5cm}\mbox{\!\!\!\includegraphics[width=.14\textwidth]{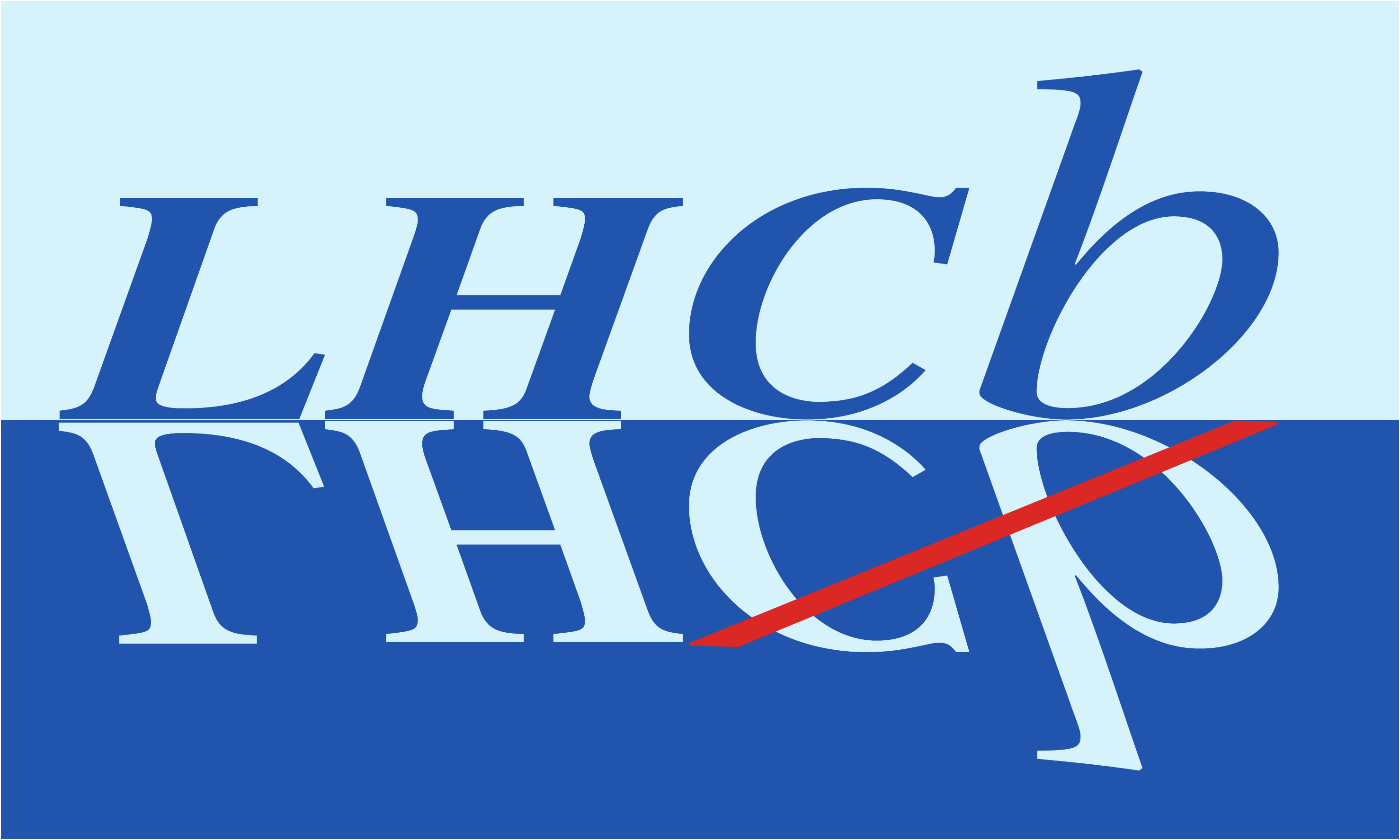}} & &}%
{\vspace*{-1.2cm}\mbox{\!\!\!\includegraphics[width=.12\textwidth]{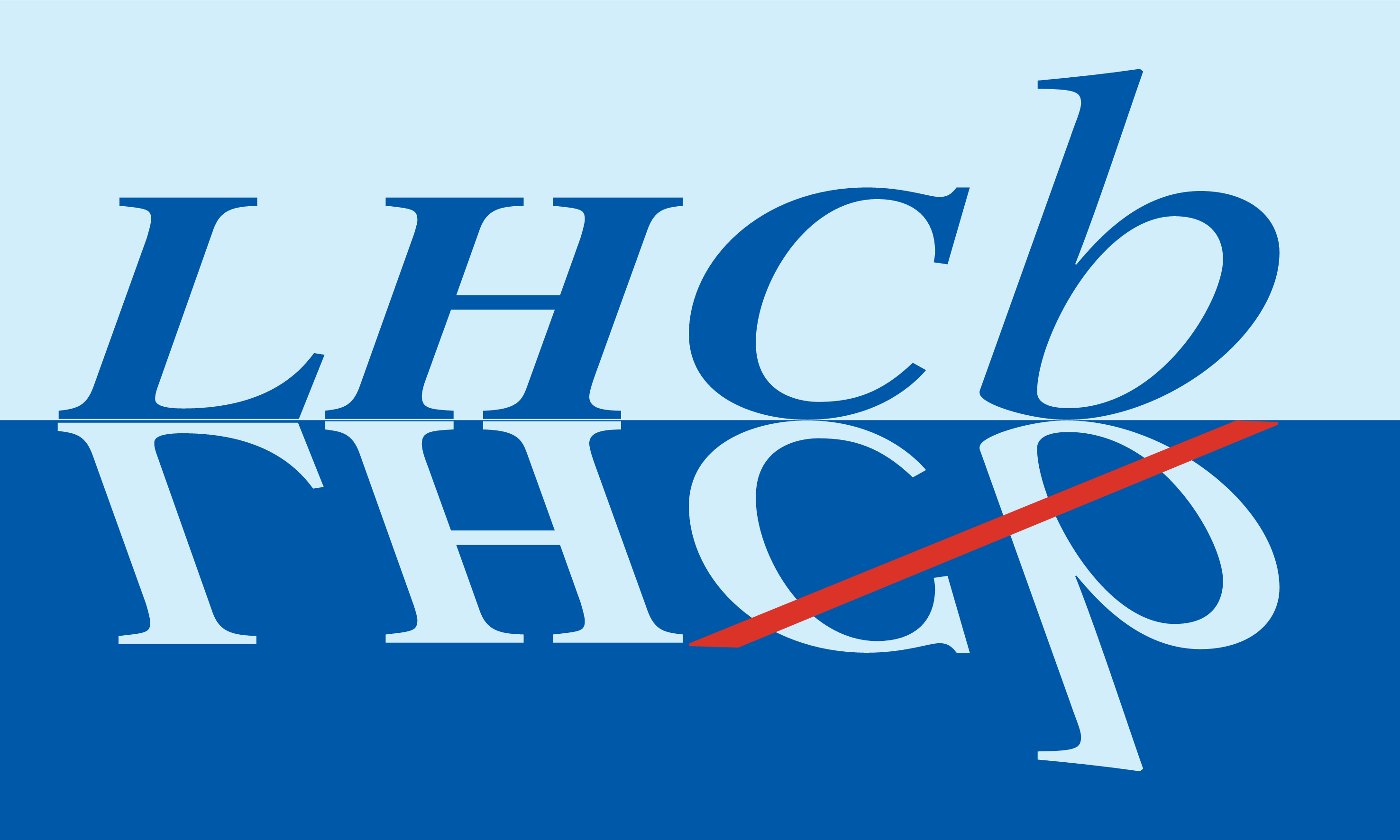}} & &}%
\\
 & & CERN-PH-EP-2012-094 \\  
 & & LHCb-PAPER-2011-045 \\  
 & & 22 January 2020 \\ 
 & & \\
\end{tabular*}

\vspace*{2cm}

{\normalfont\bfseries\boldmath\huge
\begin{center}
  \papertitle 
\end{center}
}

\vspace*{0.5cm}

\begin{center}
\paperauthors\footnote{Authors are listed at the end of this paper.}
\end{center}

\vspace{\fill}

\begin{abstract}
  \noindent
The differential cross-section for the inclusive production
of $\psi(2S)$ mesons in $pp$ collisions at $\sqrt{s}$=7\tev 
has been measured with the LHCb detector. The data sample corresponds
to an integrated luminosity of $36$\invpb. 
The $\psi(2S)$ mesons are reconstructed in the decay channels 
$\psi(2S) \rightarrow \mu^+ \mu^-$ and
$\psi(2S) \rightarrow \jpsi \pi^+ \pi^-$, with the 
$\jpsi$ meson decaying into two muons.
Results are presented both for promptly produced $\psi(2S)$ mesons 
and for those originating from $b$-hadron decays. In the kinematic range 
$p_{\rm T}(\psi(2S)) \le 16$\gevc and $2 < y(\psi(2S)) \le 4.5$ we measure
\begin{eqnarray*}
\sigma_{\rm prompt}(\psi(2S)) &=& 1.44 \pm 0.01~(\text{stat})\pm 0.12~(\text{syst})^{+0.20}_{-0.40}~(\text{pol})~{\rm \upmu b}, \\
\sigma_{b}(\psi(2S)) &=& 0.25 \pm 0.01~(\text{stat}) \pm 0.02~(\text{syst})~{\rm \upmu b}, 
\end{eqnarray*}
where the last uncertainty on the prompt cross-section is due to 
the unknown $\psi(2S)$ polarization. 
Recent QCD calculations are found to be in good agreement
with our measurements.
Combining the present result with the LHCb $\jpsi$ measurements we determine the inclusive branching fraction
\begin{equation*}
\mathcal{B}(b \rightarrow \psi(2S) X) = (2.73 \pm 0.06~(\text{stat}) \pm 0.16~(\text{syst}) \pm 0.24~(\text{BF})) 
\times 10^{-3},
\end{equation*}
where the last uncertainty is due to the $\mathcal{B}(b \rightarrow J/\psi X)$,
$\mathcal{B}(J/\psi \rightarrow \mu^+ \mu^-)$
and $\mathcal{B}(\psi(2S) \rightarrow e^+ e^-)$ branching fraction uncertainties.

\textbf{All above results are corrected by an erratum included as an \mbox{appendix}.}
  
\end{abstract}

\vspace*{0.5cm}

\begin{center}
  Published in  Eur.~Phys.~J.~C 72 (2012) 2100; Eur.~Phys.~J.~C 80 (2020) 49
\end{center}

\vspace{\fill}

{\footnotesize 
\centerline{\copyright~\papercopyright. \href{\paperlicenceurl}{\paperlicence}.}}
\vspace*{2mm}

\end{titlepage}


\newpage
\setcounter{page}{2}
\mbox{~}
%
%
%
%

\cleardoublepage


\renewcommand{\thefootnote}{\arabic{footnote}}
\setcounter{footnote}{0}



\pagestyle{plain} 
\setcounter{page}{1}
\pagenumbering{arabic}


%

\newcommand{\lumin}[2]{{#1}\;10^{#2}\;\mathrm{cm^{-2} s^{-1}}}
\newcommand{\lolumi}{\mbox{$\lum\ =\,2 \cdot 10^{32}$~cm$^{-2}$ s$^{-1}$~}}
\newcommand{\hilumi}{\mbox{$\lum\ =\,5 \cdot 10^{32}$~cm$^{-2}$ s$^{-1}$~}}
\long\def\symbolfootnote[#1]#2{\begingroup%
\def\thefootnote{\fnsymbol{footnote}}\footnote[#1]{#2}\endgroup}
\def\jpsi{\ensuremath{J\mskip -3mu/\mskip -2mu\psi\mskip 2mu}}
\newcommand{\tomm}{\ensuremath{\psi(2S) \to \mu^+ \mu^-}\xspace}
\newcommand{\toee}{\ensuremath{\psi(2S) \to e^+ e^-}\xspace}  
\newcommand{\topp}{\ensuremath{\psi(2S) \to \jpsi \pi^+ \pi^-}\xspace}  
\newcommand{\toJmm}{\ensuremath{\jpsi \to \mu^+ \mu^-}\xspace}
\newcommand{\tonmm}{\ensuremath{\psi(2S) \to \mu \mu}\xspace}
\newcommand{\tonee}{\ensuremath{\psi(2S) \to e e}\xspace}  
\newcommand{\tonpp}{\ensuremath{\psi(2S) \to \jpsi \pi \pi}\xspace}  
\newcommand{\toJnmm}{\ensuremath{\jpsi \to \mu \mu}\xspace}
\newcommand{\Bmm}{\ensuremath{\mathcal{B}(\psi(2S) \to \mu^+ \mu^-)}\xspace}
\newcommand{\Bee}{\ensuremath{\mathcal{B}(\psi(2S) \to e^+ e^-)}\xspace}  
\newcommand{\Bpp}{\ensuremath{\mathcal{B}(\psi(2S) \to \jpsi \pi^+ \pi^-)}\xspace}  
\newcommand{\BJmm}{\ensuremath{\mathcal{B}(\jpsi   \to \mu^+ \mu^-)}\xspace}
\newcommand{\Bnmm}{\ensuremath{\mathcal{B}(\psi(2S) \to \mu \mu)}\xspace}
\newcommand{\Bnee}{\ensuremath{\mathcal{B}(\psi(2S) \to e e)}\xspace}  
\newcommand{\Bnpp}{\ensuremath{\mathcal{B}(\psi(2S) \to \jpsi \pi \pi)}\xspace}  
\newcommand{\BnJmm}{\ensuremath{\mathcal{B}(\jpsi \to \mu \mu)}\xspace}
\newcommand\chidof{\ensuremath{\chi^2/\text{ndf}}}

\newcommand\Tstrut{\rule{0pt}{2.6ex}}   
\newcommand\Bstrut{\rule[-1.2ex]{0pt}{0pt}}

\section{Introduction}
Since its discovery, heavy quarkonium  
has been one of the most important test laboratories for the development
of QCD at the border between the perturbative and non-perturbative regimes,
resulting in the formulation of the
nonrelativistic QCD (NRQCD) factorisation formalism~\cite{bib:Caswell,bib:Bodwin}. 
However, prompt production studies  
carried out at the Tevatron collider in the early 1990s~\cite{bib:cdf1}
made clear that NRQCD calculations, 
based on the leading-order (LO)  colour-singlet model (CSM), 
failed to describe the absolute value 
and the transverse momentum (\pt) dependence of the charmonium production cross-section and
polarization data.   
Subsequently, the inclusion of colour-octet amplitudes in the 
NRQCD model has reduced
the discrepancy between theory and experiment, albeit at the price
of tuning $ad~hoc$ some matrix elements~\cite{bib:Bodwin}. On the other hand, 
recent computations of the next-to-leading-order (NLO) and 
next-to-next-to-leading-order (NNLO) terms 
in the CSM yielded predictions in better 
agreement with experimental data, thus resurrecting interest in the colour-singlet
framework.   
Other models have been proposed and it is important  
to test them in the LHC energy regime~\cite{HQP,HQ}.

Heavy quarkonium is also produced from $b$-hadron decays. It can be distinguished
from promptly produced quarkonium exploiting its finite decay time. QCD predictions 
are based on the Fixed-Order-Next-to-Leading-Log (FONLL) approximation for the $b\bar{b}$
production cross-section. The FONLL approach improves NLO results by resumming $p_{\text{T}}$ 
logarithms up to the next-to-leading order~\cite{bib:FONLL_1,bib:FONLL_2}.

To allow a comparison with theory,
promptly produced quarkonia should  
be separated from those coming from $b$-hadron decays and 
from those cascading from higher mass states (feed-down).
The latter contribution strongly affects $\jpsi$ production and 
complicates the interpretation of prompt  
$\jpsi$ data. On the other hand, $\psi(2S)$ charmonium has no
appreciable feed-down from higher mass states and therefore the results can be
directly compared with the theoretical predictions, making it an ideal laboratory
for QCD studies.  

This paper presents a measurement of the \psitwos meson 
production cross-section in $pp$ collisions at the
centre-of-mass energy $\sqrt{s}$ = 7\tev. The data were collected by the LHCb
experiment in 2010 and correspond to an integrated luminosity of
35.9$\pm$1.3\invpb. The analysis is similar to that described in Ref.~\cite{bib:jpsi} 
for the \jpsi production studies; in particular, the separation between promptly 
produced $\psi(2S)$ and those originating from $b$-hadron decays is based on the
reconstructed decay vertex information.  
Two decay modes of the \psitwos meson have been used: $\tomm$ and
$\topp$ followed by $\toJmm$. The $\jpsi \pi^+ \pi^-$ mode,
despite a larger background and a lower reconstruction efficiency, is used 
to cross-check and average the results, and to extend the accessible phase space.
The production of \psitwos meson at the LHC has also been studied 
at the CMS experiment~\cite{bib:cms}. 

\section{The LHCb detector and data sample}
\label{sec:detector}
The LHCb detector is a forward spectrometer~\cite{detectorpaper}, 
designed for precision studies of \CP violation and rare decays of $b$- and $c$-hadrons. 
Its tracking acceptance covers approximately the pseudorapidity
region $2 < \eta < 5$.   
The detector elements are placed along the beam line of the LHC starting with the vertex detector, 
a silicon strip device that surrounds the $pp$ interaction region and is positioned at 8\mm
from the beams during collisions. It provides precise measurements of the positions of the primary
$pp$ interaction vertices and decay vertices of long-lived hadrons, and contributes to the measurement
of particle momenta. Other detectors used for momentum measurement include a large area silicon
strip detector located before a dipole magnet of approximately 4\,Tm, and a combination of silicon strip detectors 
and straw drift chambers placed downstream. Two ring imaging Cherenkov detectors are used to
identify charged hadrons. Further downstream an electromagnetic calorimeter is used
for photon and electron detection, followed by a hadron calorimeter.
The muon detection consists of five muon stations equipped with multi-wire proportional 
chambers, with the exception of the centre of the first station using triple-GEM detectors.

The LHCb trigger system consists of a hardware level, based on information from the
calorimeter and the muon systems and designed to reduce the frequency of accepted events 
to a maximum of 1\mhz, followed by a software level which applies a full event reconstruction.
In the first stage of the software trigger a partial event reconstruction is performed. 
The second stage performs a full event reconstruction to further enhance the signal purity.

The analysis uses events selected by single muon or dimuon triggers. 
The hardware trigger requires one muon candidate with a \pt larger than 
1.4\gevc or two muon candidates with a \pt larger than 560\mevc and 480\mevc.
In the first stage of the software trigger, either of the two following
selections is required. The first selection confirms the single muon trigger candidate and applies a 
harder cut on the muon \pt at 1.8\gevc. The second selection confirms the dimuon trigger  
candidate by requiring the opposite charge of the two muons 
and adds a requirement to the dimuon mass to be greater than 2.5\gevcc. 
In the second stage of the software trigger, two selections are used for the \tomm mode. The first
tightens the requirement on the dimuon mass to be greater than 2.9\gevcc and it applies to the firtst
8\invpb of the data sample. Since this selection was subsequently prescaled by a factor five, for the 
largest fraction of the remaining data (28\invpb) a different selection is used, which in addition 
requires a good quality primary vertex and tracks for the dimuon system. For the $\jpsi\pi^+ \pi^-$ mode
only one selection is used which requires the combined dimuon mass to be in a $\pm$120\mevcc 
mass window around the nominal $\jpsi$ mass. 
To avoid that a few events with high occupancy dominate the software trigger CPU time,
a set of global event cuts is applied on the hit multiplicity of each subdetector used by
the pattern recognition algorithms, effectively rejecting events with a large number
of pile-up interactions.

The simulation samples used for this analysis are based on the \pythia 6.4 generator~\cite{bib:pythia}
configured with the parameters detailed in Ref.~\cite{bib:pythiaparam}. 
The prompt charmonium production processes activated in \pythia are
those from the leading-order colour-singlet and colour-octet mechanisms. Their implementation
and the parameters used are described in detail in Ref.~\cite{bib:vagnoni}.
The \evtgen package~\cite{bib:evtgen}
is used to generate hadron decays and the \geant package~\cite{bib:geant} for the detector simulation.
The QED radiative corrections to the decays are generated using the \photos package~\cite{bib:photos}.

\section{Signal yield}
\label{sec:selection}
The two modes, \tomm and \topp, have different decay and 
background characteristics, therefore dedicated selection criteria 
have been adopted. 
The optimisation of the cuts has been performed using the
simulation. A common requirement is that the tracks,
reconstructed in the full tracking system and passing the trigger requirements,
must be of good quality ($\chidof <4$, where ndf is the number of degrees of freedom)
and share the same vertex with fit probability $P(\chi^2) > 0.5 \%$ ($\tomm$) and 
$P(\chi^2) > 5 \%$ ($\topp$). 
A cut $\pt >$ 1.2\gevc is applied for the muons from the $\tomm$ decay.
For muons from $\jpsi(\mu^+ \mu^-) \pi^+ \pi^-$ 
we require a momentum larger than 8\gevc and $\pt >$ 0.7\gevc. Finally the
rapidity of the reconstructed \psitwos is required to satisfy 
the requirement $2 < y \le 4.5$.  
 
\begin{figure}[htbp]
\begin{center}
{\includegraphics*[width=0.45\textwidth]{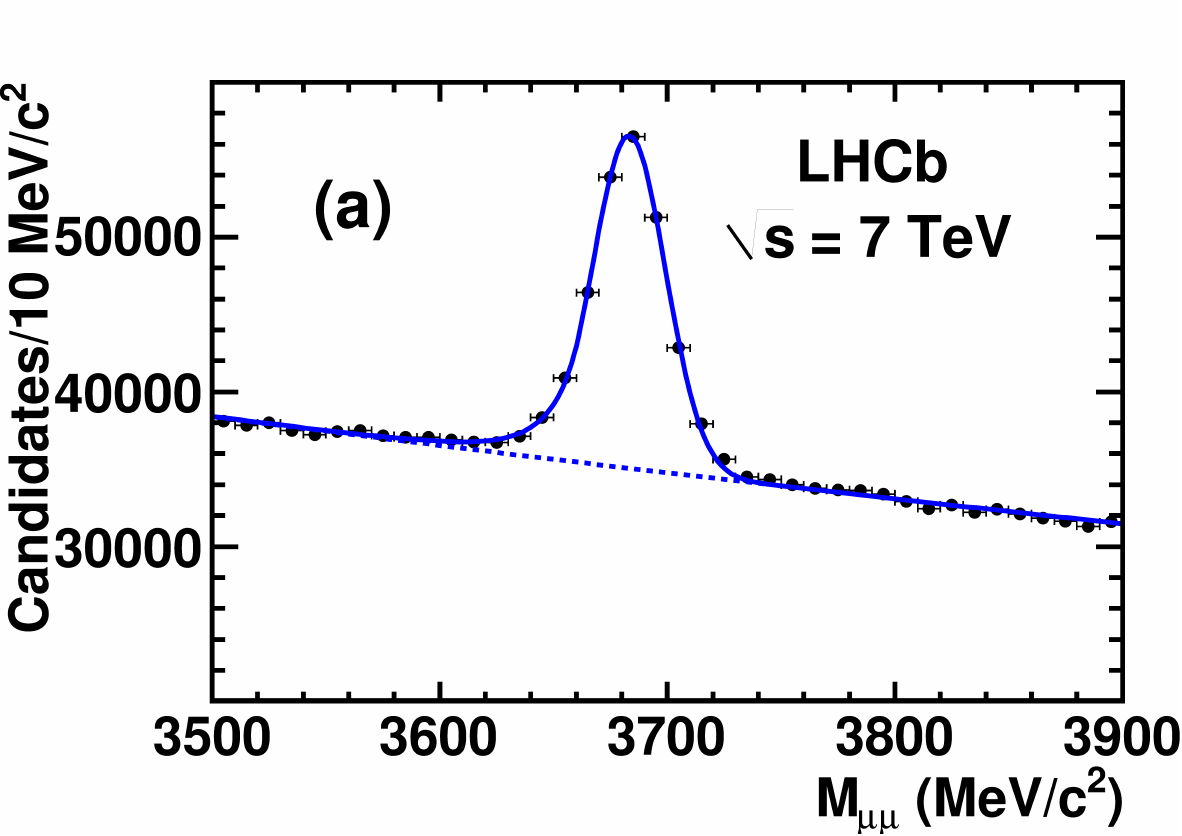}}\qquad
{\includegraphics*[width=0.45\textwidth]{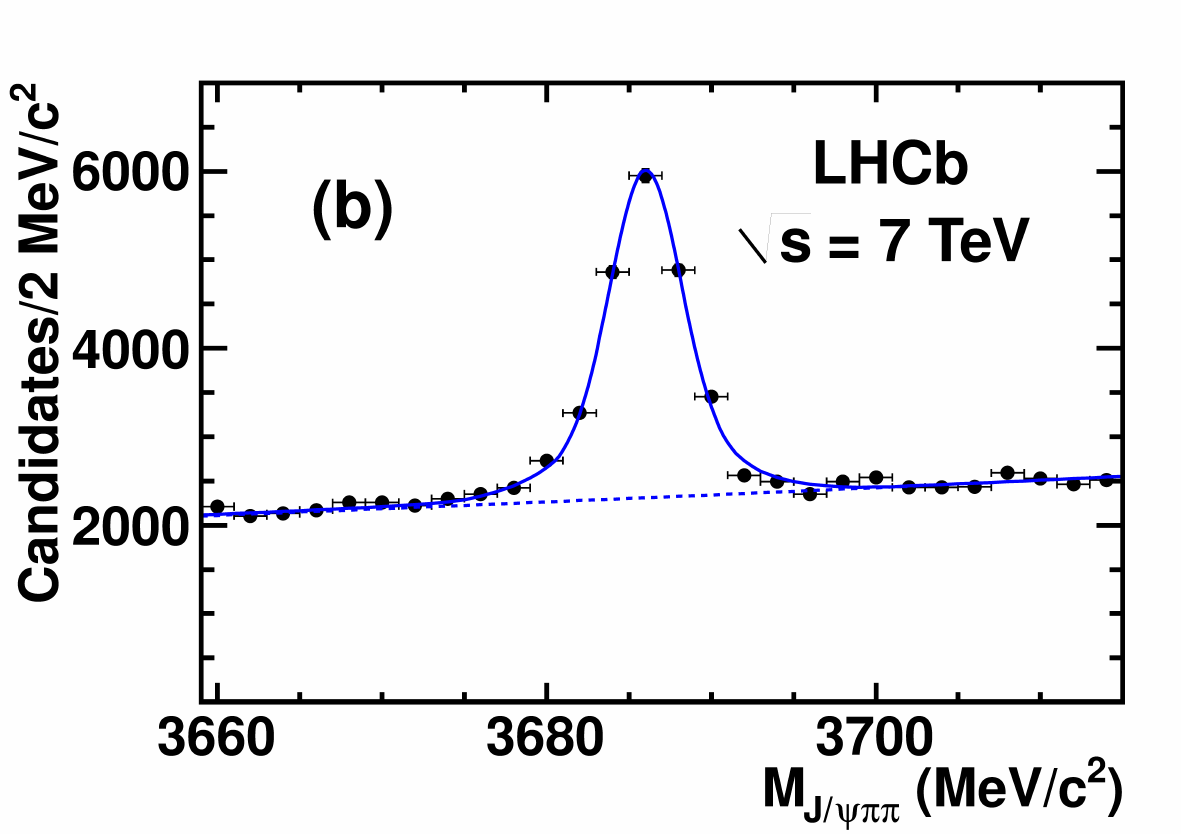}}
\end{center}
\caption{Invariant mass distribution for all $\psi(2S)$ candidates 
passing the selection cuts for the $\mu^+ \mu^-$ decay (a) and 
the $\jpsi(\mu^+ \mu^-) \pi^+ \pi^-$ decay (b).} 
\label{fig:overall_MM}
\end{figure}

The $\tomm$ invariant mass spectrum for all selected candidates
is shown in Fig.~\ref{fig:overall_MM}(a). The fitting function 
is a Crystal Ball~\cite{bib:cb2} describing the signal 
plus an exponential function for the background. In total 90600$\pm$690 
signal candidates are found in the \pt range 0--12\gevc. 
The mass resolution is 16.01$\pm$0.12\mevcc and the Crystal Ball parameters
that account for the radiative tail are obtained from the simulation.

For the $\psitwos \to \jpsi(\mu^+ \mu^-) \pi^+ \pi^-$ decay, 
both pions are required to have $\pt >$~0.3\gevc and the sum of the two-pion 
transverse momenta is required to be larger than 0.8\gevc.
The quantity $Q = M(\jpsi\pi^+\pi^-) - M(\pi^+\pi^-) - M(\mu^+\mu^-)$ is required to be 
$\le$ 200\mevcc and to improve the mass resolution the dimuon invariant mass 
$M_{\mu^+ \mu^-}$ is constrained in the fit to the nominal $\jpsi$ mass value~\cite{bib:PDG}.
Finally, both $\jpsi$ and $\psitwos$ candidates must have $\pt >$ 2\gevc. 
The invariant mass spectrum is shown in Fig.~\ref{fig:overall_MM}(b) for all
selected candidates.  
For this decay mode the peak is described by the sum of two Crystal Ball functions for the
signal plus an exponential function for the background. The number
of signal candidates is 12300$\pm$200, the mass resolution is
2.10$\pm$0.07\mevcc, and the Crystal Ball tail parameters 
are fixed to the values obtained from the simulation.

The fits are repeated in each \psitwos \pt bin to obtain the number of signal 
and background candidates for both decays.  

\section{Cross-section measurement}
\label{xsec}

The differential cross-section for the inclusive \psitwos meson 
production is computed from 
\begin{equation}
\frac{d\sigma}{d\pt}(\pt)= \frac{N_\text{sig}(\pt)}{\lum~\epsilon_{\rm tot}(\pt)~\BR~\Delta\pt} 
\label{eq:2diffXsec}
\end{equation}
where $d\sigma/d\pt$ is the average cross-section in the given \pt bin,
integrated over the rapidity range $2 < y \le 4.5$,  
$N_\text{sig}(\pt)$ is the number of signal candidates determined from the mass fit
for the decay under study, $\epsilon_{\rm tot}(\pt)$ is the total detection efficiency
including acceptance and trigger effects, $\BR$ denotes the relevant branching fraction
and $\Delta\pt$ is the bin size. All branching fractions are taken from Ref.~\cite{bib:PDG}:  
\Bee = $(7.72 \pm 0.17) \times 10^{-3}$, \Bpp = $(33.6 \pm 0.4) \times 10^{-2}$ and 
\BJmm = $(5.93 \pm 0.06) \times 10^{-2}$.
Assuming lepton universality, we use the dielectron branching fraction \Bee in Eq.~(\ref{eq:2diffXsec}),
since \Bmm is less precisely known. 
$\lum$ is the integrated luminosity, which is calibrated using both Van der Meer scans~\cite{bib:vdm,bib:vdm2}
and a beam-profile method~\cite{bib:massi}. A detailed description of the two methods is given in Ref.~\cite{bib:lumi}. 
The knowledge of the absolute luminosity scale is used to calibrate 
the number of tracks in the vertex detector, which is found to be stable throughout the data taking 
period and can therefore be used to monitor the instantaneous luminosity of the entire data sample. 
The integrated luminosity of the data sample used in this analysis is determined to be 35.9\invpb. 

The total efficiency, $\epsilon_{\rm tot}(\pt)$, is a product of three contributions: 
the geometrical acceptance, the combined detection, reconstruction and selection efficiency, 
and the trigger efficiency. Each contribution has been determined using simulated events 
for the two decay channels. In order to evaluate the trigger efficiency, the trigger selection 
algorithms used during data taking are applied to the simulation.

\begin{figure}[htb]
\begin{center}
{\includegraphics*[width=0.45\textwidth]{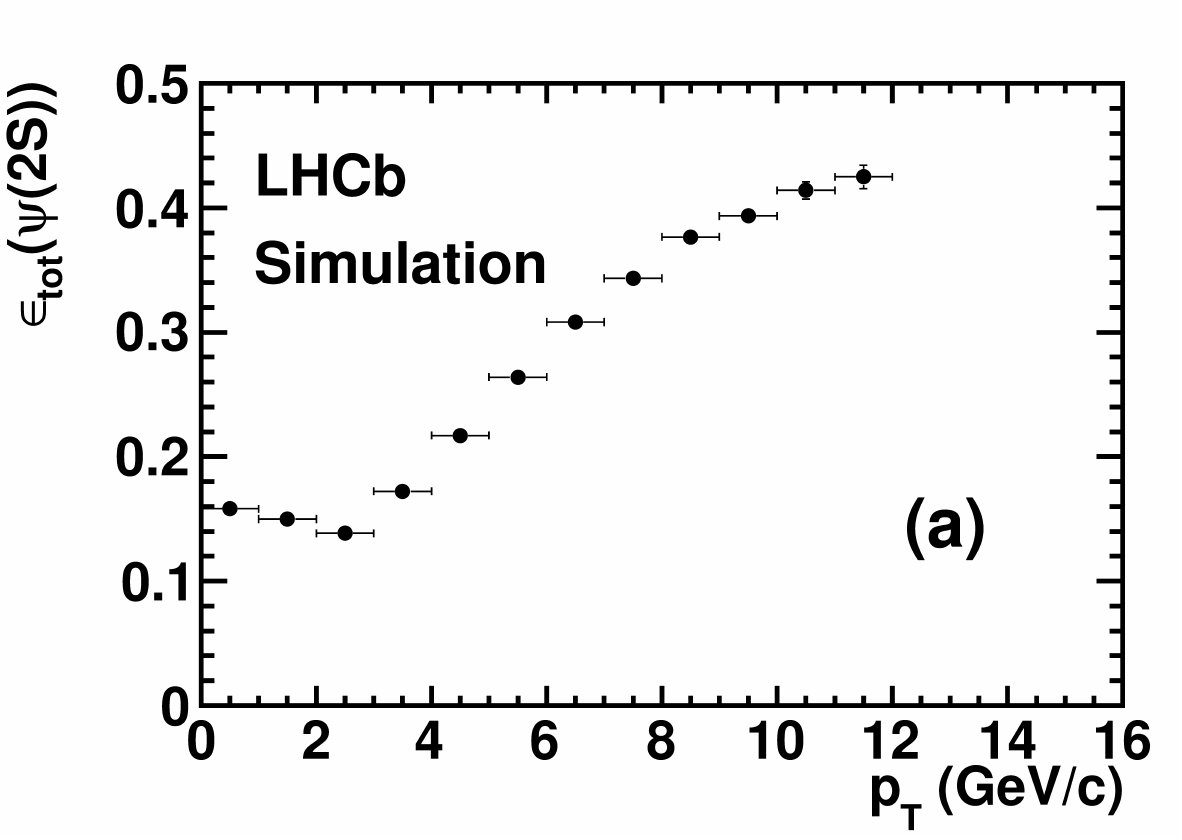}}\qquad
{\includegraphics*[width=0.45\textwidth]{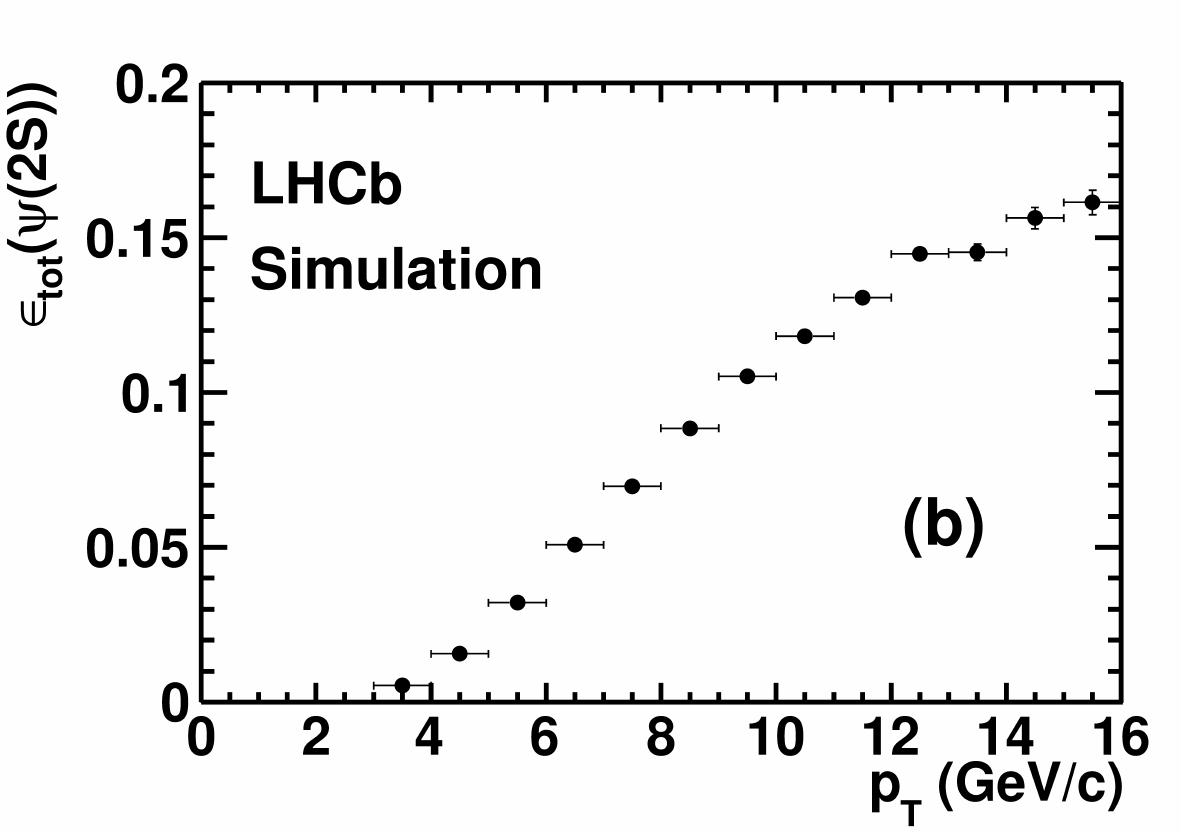}}
\caption{Total efficiency {\em vs.} \pt computed from simulation for
unpolarized \psitwos mesons for $\psi(2S) \to \mu^+\mu^-$ (a) and
$\psi(2S) \to \jpsi(\mu^+ \mu^-) \pi^+ \pi^-$ (b).}
\label{fig:psi2S_tot}
\end{center}
\end{figure}

The total efficiency {\em vs.} \pt for the two channels, assuming the $\psi(2S)$ meson unpolarized,
is shown in Fig.~\ref{fig:psi2S_tot}. 
Extensive studies on dimuon decays of prompt $\jpsi$~\cite{bib:jpsi}, $\psi(2S)$ 
and $\Upsilon$~\cite{bib:giulia} mesons have shown that the total efficiency in the LHCb detector depends strongly 
on the initial polarization state of 
the vector meson. This effect is absent for \psitwos mesons coming from $b$-hadron decays.
In fact for these events the natural polarization axis is the \psitwos meson flight direction
in the $b$-hadron rest frame, while the \psitwos meson appears unpolarized along  
its flight direction in the laboratory. Simulations~\cite{bib:jpsi} and  
measurements from CDF \cite{bib:cdf} confirm this. 
We do not measure the \psitwos meson polarization but we assign 
a systematic uncertainty to the unpolarized efficiencies 
in the case of prompt production.    
Events are generated with polarizations corresponding to the two extreme cases of fully
transverse or fully longitudinal polarization and the efficiency is re-evaluated. The difference between 
these results and those with the unpolarized sample is taken as an estimate of the systematic 
uncertainty.

A similar effect exists for the $\jpsi$ meson emitted in the  
$\psi(2S) \to \jpsi(\mu^+ \mu^-) \pi^+ \pi^-$ decay.
However, in this case,
the $\psi(2S)$ meson polarization is fully transferred to the $\jpsi$ meson
since, as measured by the BES collaboration~\cite{bes}, 
the two pions are predominantly in the $S$-wave configuration\footnote[1]{The small fraction of 
$D$-wave measured in Ref.~\cite{bes} has a negligible impact on our conclusion.} and the dipion-$\jpsi$ 
system is also in a $S$-wave configuration. 
This has been verified with data and is correctly reproduced by the simulation. 
Therefore the systematics due to polarization are fully correlated 
between the two channels and we use the systematic  uncertainties 
computed for \tomm also for the \topp decay. 

In order to separate prompt $\psi(2S)$ mesons from those produced 
in $b$-hadron decays, we use the pseudo-decay-time variable defined as 
$t=d_z (M/p_z)$, where $d_z$ is the separation along the beam axis between 
the $\psi(2S)$ decay vertex and the primary vertex, $M$ is the nominal mass 
of the $\psi(2S)$ and $p_z$ is the component of its momentum along the beam axis.
In case of multiple primary vertices reconstructed
in the same event, that which minimises $|d_z|$ has been chosen. 
The prompt component is distributed as a Gaussian function around $t=0$, 
with width corresponding to the experimental resolution, while for the $\psi(2S)$ 
from $b$-hadron decays the $t$ variable is distributed according to an approximately 
exponential decay law, smeared in the fit with the experimental resolution.
The choice of taking the primary vertex which minimises $|d_z|$ could in principle
introduce a background component in the pseudo-decay-time distribution arising 
from the association of the $\psi(2S)$ vertex to a wrong primary vertex. 
The effect of such background is found to be of the order of
0.5\% in the region around $t=0$ and has been neglected.
The function used to fit the $t$ distribution   
in each \pt bin is
\begin{equation}
F(t;f_\text{p},\sigma,\tau_b)=N_\text{sig}\left[f_\text{p}\delta(t)+
(1-f_\text{p})\theta(t)\frac{e^{-\frac{t}{\tau_b}}}{\tau_b}\right]
\otimes\frac{e^{-\frac{1}{2}(\frac{t}{\sigma})^2}}{\sqrt{2\pi}\sigma}+ 
N_\text{bkg}f_\text{bkg}(t;\boldsymbol \Theta)
\label{eq:SigFitFunc}
\end{equation}
where $N_\text{sig}$ and $N_\text{bkg}$ are respectively 
the numbers of signal and background candidates obtained from the mass fit.  
The fit parameters are the prompt fraction, $f_\text{p}$, the standard deviation
of the Gaussian resolution function, $\sigma$, and the lifetime describing the 
long-lived component of $\psi(2S)$ mesons coming from $b$-hadron decays, $\tau_b$. 
In principle, all fit parameters 
are dependent on \pt. The function $f_\text{bkg}(t;\boldsymbol \Theta)$ 
models the background component in the distribution and is defined as the sum of a
$\delta$ function and a Gaussian function for the prompt background, plus two 
exponential functions for the positive tail and one exponential function for the 
negative tail, all convolved with a Gaussian function to account for the detector resolution.
The array of parameters $\boldsymbol \Theta$ is determined from a fit to the $t$ distribution 
of the events in the mass sidebands. 

\begin{figure}[ht]
\begin{center}
{\includegraphics[width=0.45\textwidth]{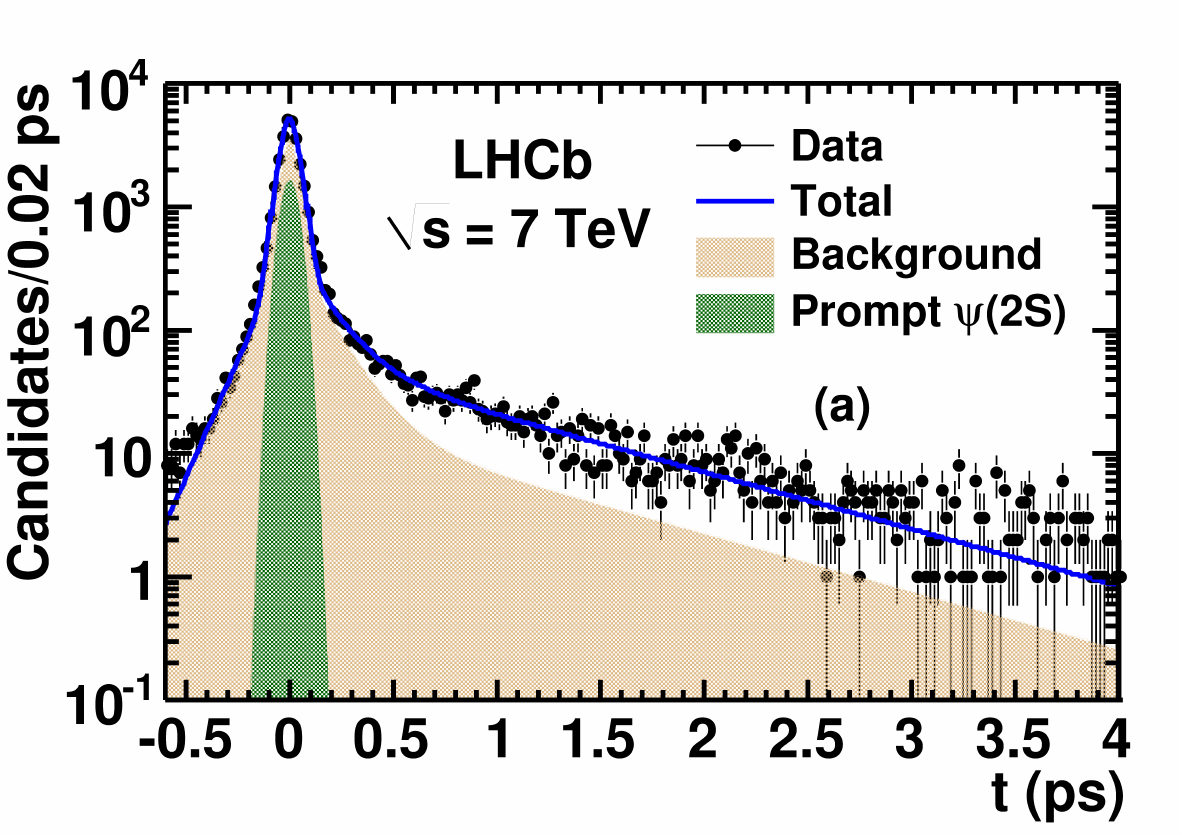}}\qquad
{\includegraphics[width=0.45\textwidth]{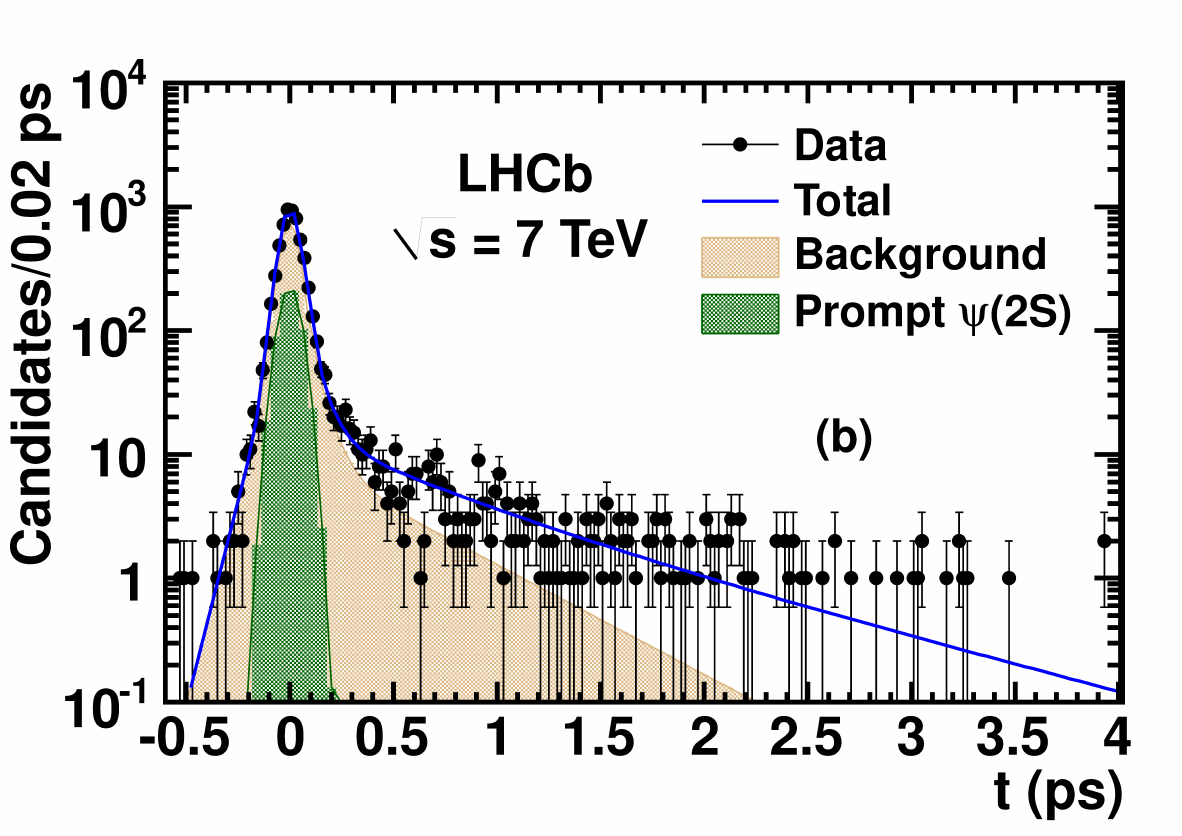}}
\caption{Pseudo-decay-time distribution for $\tomm$ (a) 
and $\topp$(b) in the \pt range $4<\pt\le5$ \gevc, 
showing the background and prompt contributions. \textbf{This figure is updated in the Erratum (Appendix ~\ref{erratum})}}
\label{fig:tauBins}
\end{center}
\end{figure}

\begin{figure}[ht]
\begin{center}
{\includegraphics*[width=0.6\textwidth]{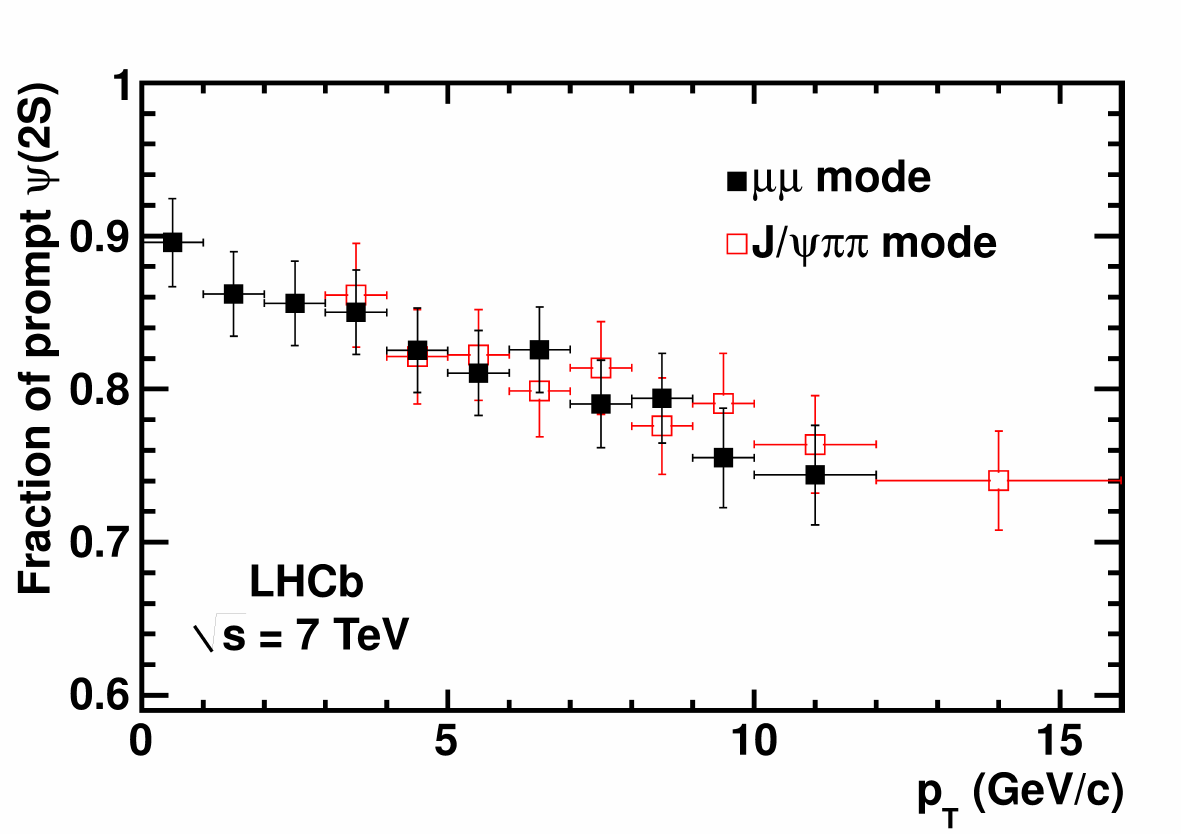}}
\caption{Fraction of prompt $\psi(2S)$ as a function of \pt for the $\mu^+ \mu^-$ 
mode (solid squares) and the $\jpsi \pi^+ \pi^-$ mode (open squares). Error bars
include the statistical uncertainties and the systematic uncertainties
due to the fitting procedure. \textbf{This figure is updated in the Erratum (Appendix ~\ref{erratum})}.}
\label{fig:prompt_frac} 
\end{center}
\end{figure}

As an example, the pseudo-decay-time distributions for $\tomm$ and $\topp$ 
in the \pt range $4<\pt\le5$\gevc are presented in Fig.~\ref{fig:tauBins}. 
The contributions of background and prompt \psitwos mesons are also shown. 
The values of the prompt fraction, $f_\text{p}$~{\em vs.} \pt 
in the rapidity range $2 < y \le 4.5$, obtained 
for the $\mu^+ \mu^-$ and the $\jpsi \pi^+ \pi^-$ modes, are in good agreement 
as shown in Fig.~\ref{fig:prompt_frac}.

\section{Systematic uncertainties on the cross-section measurement}
\label{systematics}
A variety of sources of systematic uncertainties affecting the cross-section measurement
were taken into account and are summarised in Table~\ref{tab:syst}.

\begin{table}[htb]
\begin{center}
\caption{Systematic uncertainties included in the measurement of the cross-section.
Uncertainties labelled with $a$ are correlated between the $\mu^+ \mu^-$ and 
$\jpsi \pi^+ \pi^-$ mode, while $b$ indicates a correlation
between $\psi(2S) \rightarrow \mu^+ \mu^-$ and the $\jpsi \rightarrow \mu^+ \mu^-$
uncertainties~\cite{bib:jpsi}.}
\vspace*{0.7\baselineskip}
\begin{tabular}{c|c|c}
Uncertainty source                       & $\mu^+ \mu^-$(\%)& $\jpsi \pi^+ \pi^-$(\%)  \\ \hline
Luminosity$^{a,b}$                        &   3.5       &     3.5            \\ 
Size of simulation sample                        & 0.4--2.2    &   0.6--1.0         \\
Trigger efficiency$^a$                   &   1--8      &     1--7           \\
Global event cuts$^{a,b}$                 &   2.1       &     2.1            \\
Muon identification$^{a,b}$               &   1.1       &     1.1            \\
Hadron identification                     &    --       &     0.5            \\
Track $\chi^{2}$$^{a,b}$                   &    1        &      2             \\
Tracking efficiency$^a$                  &   3.5       &     7.3            \\
Vertex fit$^b$                            &   0.8       &     1.3            \\
Unknown polarization$^a$                  &  15--26     &    15--26          \\
Mass fit function                         &   1.1       &     0.5            \\
Pseudo-decay-time fits                    &   2.7       &     2.7            \\
\Bee                                      &   2.2       &      --            \\
\Bpp                                      &    --       &     1.2            \\
\BJmm                                     &    --       &     1.0            \\
\end{tabular}
\label{tab:syst}
\end{center}
\end{table}

A thorough analysis of the luminosity scans yields consistent results for the absolute luminosity 
scale with a precision of 3.5\%~\cite{bib:lumi}, this value being assigned as a systematic uncertainty.  
The statistical uncertainties from the finite number of simulated events on the efficiencies 
are included as a source of systematic uncertainty; this uncertainty
varies from 0.4 to 2.2\% for the $\mu^+ \mu^-$ mode and from 0.6 to 1\% for the $\jpsi \pi^+ \pi^-$ mode. 
In addition, 
we assign a systematic uncertainty in order to account for the difference
between the trigger efficiency evaluated on data 
by means of an unbiased $\mu^+ \mu^-$ sample, and the trigger 
efficiency computed from the simulation. 
This results in a bin-dependent uncertainty up to 8\% for the $\mu^+ \mu^-$
mode and up to 7\% for the $\jpsi \pi^+ \pi^-$ mode. This uncertainty is fully correlated between
the two decay modes in the overlapping \pt region.  
Finally, the statistical uncertainty on the global event cuts efficiency 
(2.1\% for both modes) is taken as an additional systematic uncertainty~\cite{bib:jpsi}.

\begin{figure}[htb]
\begin{center}
{\includegraphics*[width=0.6\textwidth]{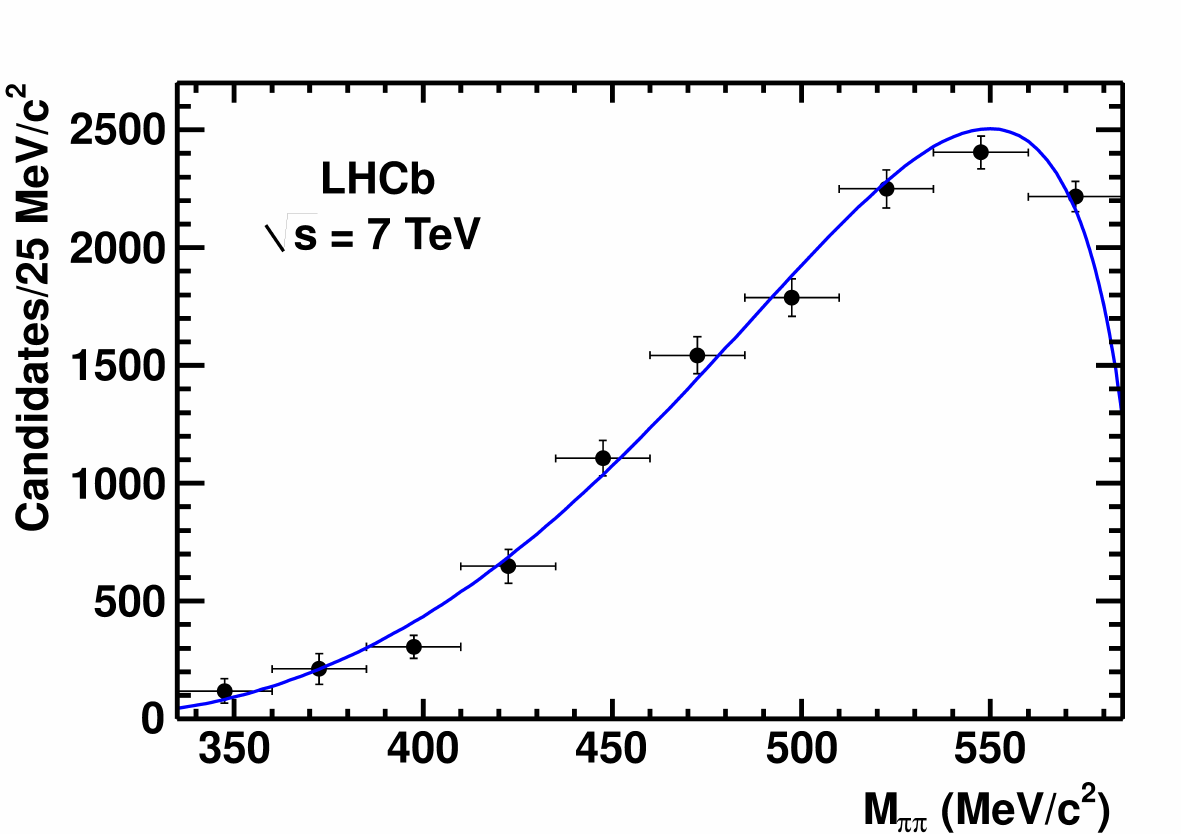}}
\caption{Dipion mass spectrum for the \topp decay. The curve shows 
the result of the fit with Eq.~(\ref{eq:dipion_mc}) corrected for the acceptance.}
\label{fig:dipion}
\end{center}
\end{figure}

To assess possible systematic differences in the acceptance 
between data and simulation for the $\jpsi \pi^+ \pi^-$ mode, we have studied
the dipion mass distribution. The LHCb simulation is based on the Voloshin-Zakharov 
model~\cite{voloshin} which uses a single phenomenological parameter $\lambda$
\begin{equation}
\frac{d\sigma}{dm_{\pi\pi}} \propto \Phi(m_{\pi \pi}) 
\left[ m_{\pi\pi}^{2} - \lambda m_{\pi}^{2} \right]^{2},
\label{eq:dipion_mc}
\end{equation} 
where $\Phi(m_{\pi \pi})$ is a phase space factor (see e.g. Ref.~\cite{bes}) 
and in the simulation $\lambda = 4$ is assumed. 
The dipion mass distribution obtained from the data is shown in Fig.~\ref{fig:dipion}.
We obtain $\lambda = 4.46 \pm 0.07 (\rm stat) \pm 0.18 (\rm syst)$,
from which we estimate a negligible systematic effect on the acceptance (0.25\%).
Our result is also in good agreement with the BES value 
$\lambda = 4.36 \pm 0.06 (\rm stat) \pm 0.17 (\rm syst)$ \cite{bes}. 

To cross-check and assign a systematic uncertainty to the determination of
the muon identification efficiency from simulation, the single track muon identification 
efficiency has been measured on data using a tag-and-probe method~\cite{bib:tagprobe}. 
This gives a correction factor for the dimuon of 1.025$\pm$0.011,
which we  apply to the simulation efficiencies. The 1.1\% uncertainty on the correction factor 
is used as systematic uncertainty.
The efficiency of the selection requirement on the dipion identification has been studied on data and simulation
and a difference of 1\% has been measured between the two. Therefore, the simulation efficiencies are
corrected for this difference and an additional systematic uncertainty of 0.5\% is included.

The $\psitwos$ selection also includes a requirement on the track fit quality. The relative difference between 
the efficiency of this requirement in simulation and data is taken as a systematic uncertainty, resulting in
an uncertainty of 0.5\% per track. 
Tracking studies show that the ratio of the track-finding efficiencies between data and simulation 
is 1.09 for the $\mu^+ \mu^-$ mode and 1.06 for the $\jpsi\pi^+ \pi^-$ mode, with an uncertainty of 3.5\% and 7.3\% 
respectively; the simulation efficiencies are corrected accordingly and the corresponding
systematic uncertainties are included.

For the requirement on the secondary vertex fit quality, a relative difference 
of 1.6\% for the $\mu^+ \mu^-$ mode and 2.6\% for the $\jpsi \pi^+ \pi^-$ mode has been measured between data and simulation. 
The simulation efficiency is therefore corrected for this  difference and
a corresponding systematic uncertainty of 0.8\% ($\mu^+ \mu^-$) and 1.3\% ($\jpsi \pi^+ \pi^-$) is assigned.

The systematic uncertainty due to the unknown polarization 
is computed as discussed in Section~\ref{xsec}.
The study done for the two extreme polarization hypotheses gives an average systematic uncertainty between 
15\% and 26\% for both modes, relative to the hypothesis of zero polarization, depending
on the \pt bin. These errors are fully correlated between the two decay modes  
and strongly asymmetric since the variations of the efficiency 
are of different magnitude for transverse and longitudinal polarizations.  

A systematic uncertainty from the fitting procedure has been estimated from the relative difference 
between the overall number of signal \psitwos and the number of signal candidates obtained by summing 
the results of the fits in the individual \pt bins. A total systematic uncertainty of 1.1\% for 
the $\mu^+ \mu^-$ mode and 0.5\% for the $\jpsi \pi^+ \pi^-$ mode is assigned.

Finally, to evaluate the systematic uncertainty on the prompt fraction from 
the \psitwos pseudo-decay-time fit we recompute $f_\text{p}$ with $\tau_b$ (see Eq.~(\ref{eq:SigFitFunc})) 
fixed to the largest and smallest value obtained in the 
\pt-bin fits. The relative variation is at most 2.7\% and this value is  assigned 
as a systematic uncertainty on $f_\text{p}$. 

\section{Cross-section results}
\label{results}
\begin{figure}[htb]
\begin{center}
{\includegraphics*[width=0.6\textwidth]{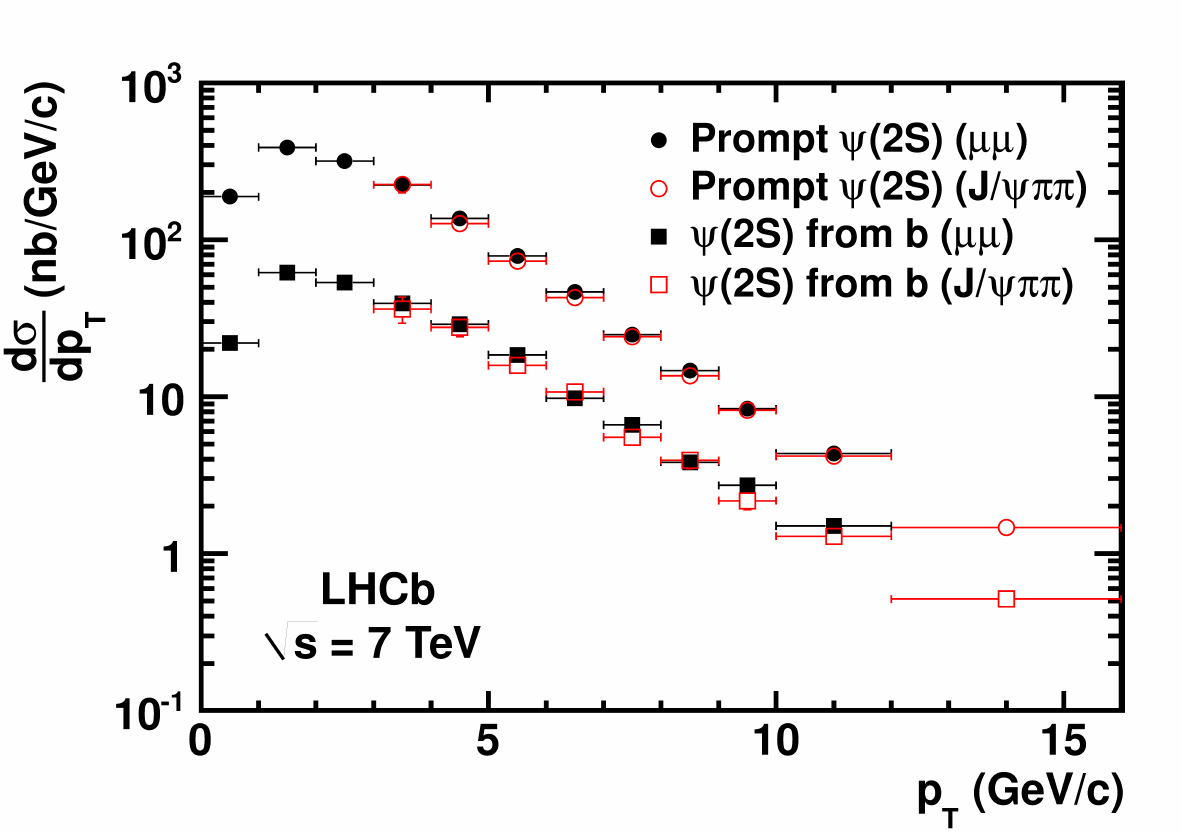}}
\caption{Comparison of the differential cross-sections measured 
for prompt $\psi(2S)$ (circles) and for $\psi(2S)$ from $b$-hadron decay (squares)
in the \tomm (solid symbols) and \topp (open symbols) modes.
Only the uncorrelated uncertainties are shown. 
\bf{This figure is obsolete. Corrected cross sections are found in Figs.~7 and 8
in the Erratum}.
} 
\label{fig:compa}
\end{center}
\end{figure}

The differential cross-sections for prompt $\psi(2S)$ and $\psi(2S)$ mesons from $b$-hadron decays
are shown in Fig.~\ref{fig:compa}, where we compare the results obtained for the \tomm and
\topp channels separately for the prompt and $b$-hadron decay components.

The values for the two cross-sections estimated using the different decay modes are consistent
within $0.5~\sigma$. A weighted average of the two measurements
is performed to extract the final result listed in Table~\ref{tab:xs}. 

\begin{table}[hbt]
\begin{center}
\caption{Cross-section values for prompt $\psi(2S)$ and $\psi(2S)$ from $b$-hadrons in different \pt bins
and in the range $2 < y \le 4.5$, evaluated as the weighted average of the $\mu^+ \mu^-$ and 
$\jpsi \pi^+ \pi^-$ channels. The first error is statistical, the second error is systematic, and the last 
asymmetric uncertainty is due to the unknown polarization of the prompt $\psi(2S)$ meson. \textbf{This table is updated in the Erratum}.}
\vspace*{0.7\baselineskip}
\begin{tabular}{c|c|c}  
\pt [GeV/$c$] & $\frac{d\sigma_{\rm prompt}}{d\pt}$ [$\frac{\rm nb}{{\rm GeV}/c}$] & 
$\frac{d\sigma_{\rm b}}{d\pt}$ [$\frac{\rm nb}{{\rm GeV}/c}$] \\ \hline
0--1\Tstrut\Bstrut & 188 $\pm$ 6 $\pm$ 18$^{+32}_{-67}$ & 22 $\pm$ 2 $\pm$ 2 \\  
1--2\Tstrut\Bstrut & 387 $\pm$ 8 $\pm$ 37$^{+60}_{-119}$ & 62 $\pm$ 3 $\pm$ 6 \\  
2--3\Tstrut\Bstrut & 317 $\pm$ 7 $\pm$ 26$^{+44}_{-88}$ & 53 $\pm$ 2 $\pm$ 4 \\  
3--4\Tstrut\Bstrut & 224 $\pm$ 6 $\pm$ 24$^{+27}_{-53}$ & 39 $\pm$ 2 $\pm$ 4 \\  
4--5\Tstrut\Bstrut & 135 $\pm$ 4 $\pm$ 13$^{+16}_{-30}$ & 29 $\pm$ 1 $\pm$ 3 \\  
5--6\Tstrut\Bstrut & 77 $\pm$ 2 $\pm$ 7$^{+9}_{-18}$ & 18 $\pm$ 1 $\pm$ 2 \\  
6--7\Tstrut\Bstrut & 46 $\pm$ 1 $\pm$ 4$^{+5}_{-10}$ & 10 $\pm$ 1 $\pm$ 1 \\  
7--8\Tstrut\Bstrut & 25 $\pm$ 1 $\pm$ 2$^{+3}_{-6}$ & 6.3 $\pm$ 0.4 $\pm$ 0.5 \\  
8--9\Tstrut\Bstrut & 14 $\pm$ 1 $\pm$ 1$^{+2}_{-3}$ & 3.9 $\pm$ 0.3 $\pm$ 0.3 \\  
9--10\Tstrut\Bstrut & 8.3 $\pm$ 0.4 $\pm$ 0.7$^{+0.9}_{-1.7}$ & 2.5 $\pm$ 0.2 $\pm$ 0.2 \\  
10--12\Tstrut\Bstrut & 4.3 $\pm$ 0.3 $\pm$ 0.4$^{+0.5}_{-0.9}$ & 1.4 $\pm$ 0.1 $\pm$ 0.1 \\  
12--16\Tstrut\Bstrut & 1.5 $\pm$ 0.1 $\pm$ 0.2$^{+0.2}_{-0.3}$ & 0.51 $\pm$ 0.04 $\pm$ 0.06 \\  
\end{tabular}
\label{tab:xs}
\end{center}
\end{table}

\begin{figure}[ht]
\begin{center}
{\includegraphics*[width=0.6\textwidth]{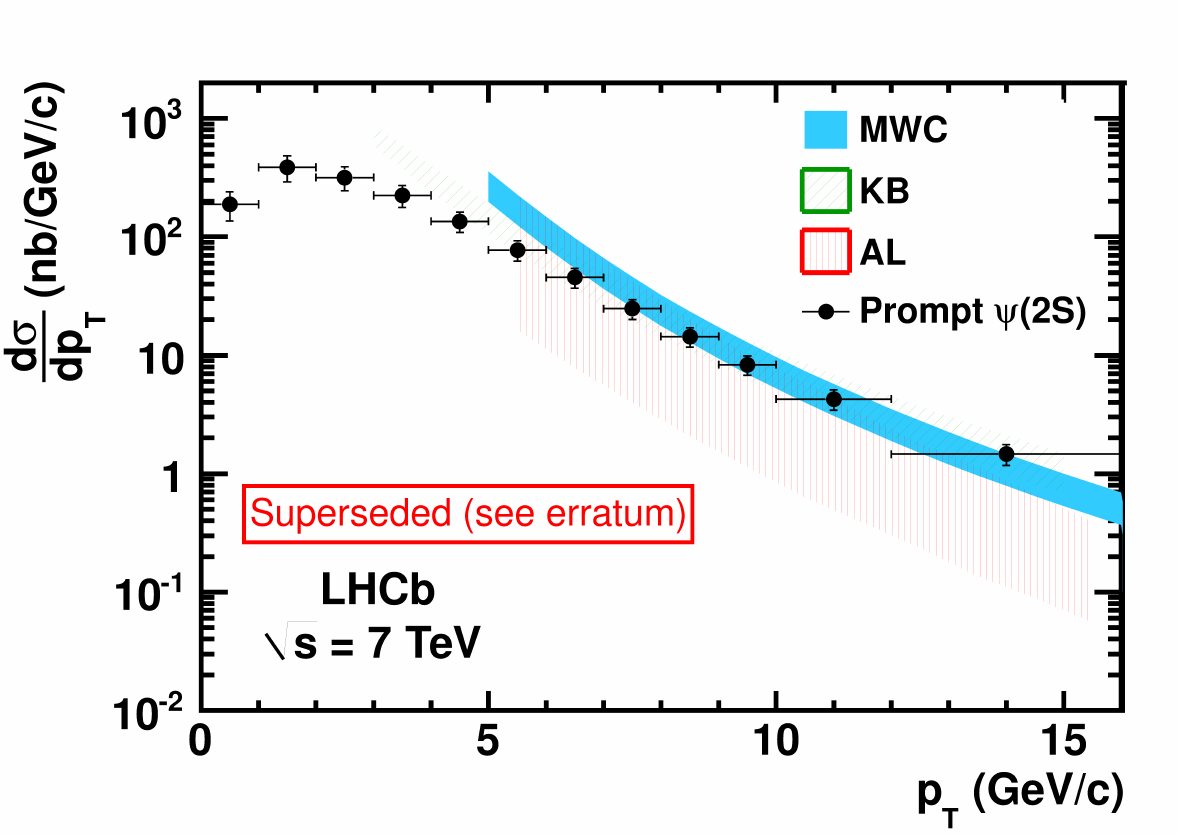}}
\caption{Differential production cross-section {\em vs.} \pt for prompt $\psi(2S)$. 
The predictions of three nonrelativistic QCD models are also shown for comparison. 
MWC~\cite{bib:chao} and KB~\cite{bib:bernd} are NLO calculations including colour-singlet 
and colour-octet contributions. AL~\cite{bib:Lansberg,bib:Lansberg2} is a colour-singlet 
model including the dominant NNLO terms. \textbf{This figure is updated in the Erratum.}} 
\label{fig:xs}
\end{center}  
\end{figure}

The differential cross-section for promptly produced $\psi(2S)$ mesons,
along with a comparison with some recent theory predictions~\cite{bib:chao,bib:bernd,bib:Lansberg,bib:Lansberg2}
tuned to the LHCb acceptance, is shown in Fig.~\ref{fig:xs}. 
In Ref.~\cite{bib:chao} and Ref.~\cite{bib:bernd} the differential prompt cross-section 
has been computed up to NLO terms in nonrelativistic QCD, including colour-singlet and 
colour-octet contributions. In Ref.~\cite{bib:Lansberg,bib:Lansberg2}
the prompt cross-section has been evaluated in a colour-singlet 
framework, including up to the dominant $\alpha_s^5$ NNLO terms. 
Experimentally the large-\pt tail behaves like $\pt^{-\beta}$ with $\beta = 4.2 \pm 0.6$ and  
is rather well reproduced, especially in the colour-octet models. 

\begin{figure}[ht]
\begin{center}
{\includegraphics*[width=0.6\textwidth]{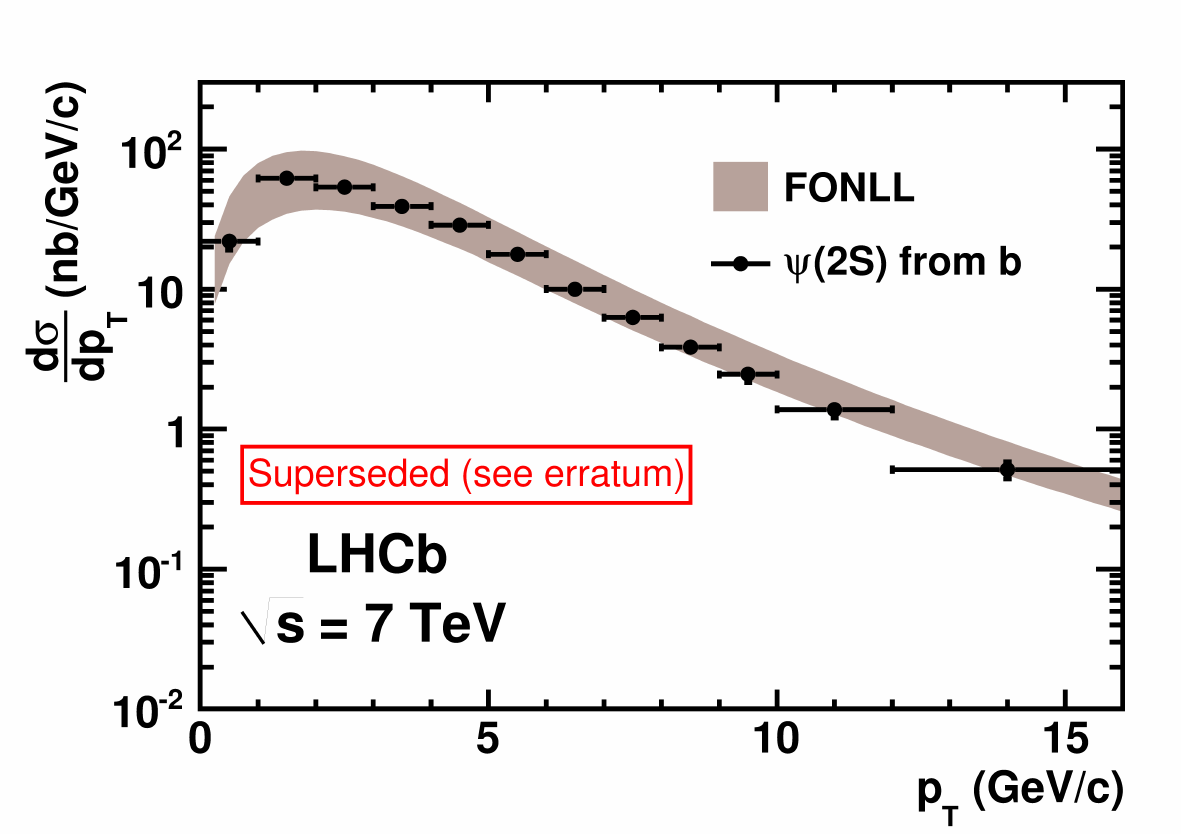}}
\caption{Differential production cross-section {\em vs.} \pt for $\psi(2S)$ from $b$-hadrons. 
The shaded band is the prediction of a FONLL calculation~\cite{bib:FONLL_1,bib:FONLL_2,bib:cacciari}. \textbf{This figure is updated in the Erratum}.} 
\label{fig:sigmab}
\end{center}  
\end{figure}

The differential cross-section for \psitwos produced in $b$-hadron decays and the comparison 
with a recent theory prediction~\cite{bib:cacciari} based on the FONLL approach~\cite{bib:FONLL_1,bib:FONLL_2} 
are presented in Fig.~\ref{fig:sigmab}. The theoretical prediction of Ref.~\cite{bib:cacciari} 
uses as input the $b \to \psi(2S) X$ branching fraction obtained in the following section. 
Experimentally the \psitwos mesons resulting from $b$-hadron decay have a slightly harder \pt 
spectrum than those produced promptly: $\beta = 3.6 \pm 0.5$. 
By integrating the differential cross-section for prompt $\psi(2S)$ and $\psi(2S)$ from $b$-hadrons 
in the range $2<y\le 4.5$ and $\pt\le$16\gevc, we obtain
\begin{align}
\sigma_{\rm prompt}(\psi(2S)) &= 1.44 \pm 0.01~(\text{stat}) \pm 0.12~(\text{syst})^{+0.20}_{-0.40}~(\text{pol})~{\rm \upmu b}, 
\nonumber \\
\sigma_{b}(\psi(2S)) &= 0.25 \pm 0.01~(\text{stat}) \pm 0.02~(\text{syst})~{\rm \upmu b}, \nonumber
\label{eq:sigma}
\end{align}
where the systematic uncertainty includes all the sources listed in 
Table~\ref{tab:syst}, except for the polarization, while the last 
asymmetric uncertainty is due to the effect of the unknown $\psi(2S)$ 
polarization and applies only to the prompt component.

\section{Inclusive $\boldsymbol{b \rightarrow \psi(2S)X}$ branching fraction measurement}
\label{sec:br}
The inclusive branching fraction for a $b$-hadron decaying to \psitwos
is presently known with 50\% precision:
$\mathcal{B}(b \to \psi(2S) X)$ = (4.8 $\pm$ 2.4) $\times 10^{-3}$~\cite{bib:PDG}.
Combining the present result for $\sigma_{b}(\psi(2S))$ with the previous measurement 
of $\sigma_{b}(\jpsi)$ \cite{bib:jpsi} we can obtain an improved value of the aforementioned 
branching fraction. 
To achieve this, it is necessary
to extrapolate the two measurements to the full phase space. The extrapolation factors for the two
decays have been determined using the LHCb simulation~\cite{bib:pythiaparam} and they have been found 
to be $\alpha_{4\pi}(J/\psi)$=5.88~\cite{bib:jpsi} and $\alpha_{4\pi}(\psi(2S))$=5.48. 
Most of the theoretical uncertainties are expected to cancel  
in the ratio of the two factors $\xi = \alpha_{4 \pi}(\psi(2S))/\alpha_{4 \pi}(J/\psi) = 0.932$, 
which is used in Eq.~(\ref{eq:br}). A systematic uncertainty 
of $3.4 \%$ is estimated for this correction and included in the final result below. 
Therefore
\begin{equation}
\frac{\mathcal{B}(b \rightarrow \psi(2S) X)}{\mathcal{B}(b \rightarrow \jpsi X)} = 
\xi \frac{\sigma_{b}(\psi(2S))}{\sigma_{b}(\jpsi)}.
\label{eq:br}
\end{equation} 

For $\sigma_{b}(\jpsi)$ we rescale the value 
in~\cite{bib:jpsi} for the new determination of the integrated 
luminosity ($\mathcal{L}$ = 5.49 $\pm$ 0.19\invpb). 
For $\sigma_{b}(\psi(2S))$ we use only the data from the $\psi(2S)~\rightarrow~\mu^+\mu^-$ mode to cancel 
most of the systematic uncertainties in the ratio. Effects due to polarization are negligible for 
mesons resulting from $b$-hadron decay. We obtain
\begin{equation*}
\frac{\mathcal{B}(b \to \psi(2S) X)}{\mathcal{B}(b \to \jpsi X)} = 0.235 \pm 0.005~(\text{stat})\pm 0.015~(\text{syst}),
\label{eq:br2}
\end{equation*} 
where the correlated uncertainties (Table~\ref{tab:syst})
between the two
cross-sections are excluded. 
By inserting the value 
$\mathcal{B}(b~\rightarrow~\jpsi X)=(1.16 \pm 0.10)\times 10^{-2}$ \cite{bib:PDG} 
we get
\begin{equation*}
\mathcal{B}(b \to \psi(2S)X) = (2.73 \pm 0.06~(\text{stat}) \pm 0.16~(\text{syst}) \pm 0.24~(\text{BF})) \times 10^{-3},
\label{eq:br3}
\end{equation*} 
where the last uncertainty originates from the uncertainty of the branching fractions   
$\mathcal{B}(b \to \jpsi X)$, $\mathcal{B}(\toee)$ and $\mathcal{B}(\toJmm)$. 

\begin{figure}[htb]
\begin{center}
{\includegraphics*[width=0.7\textwidth]{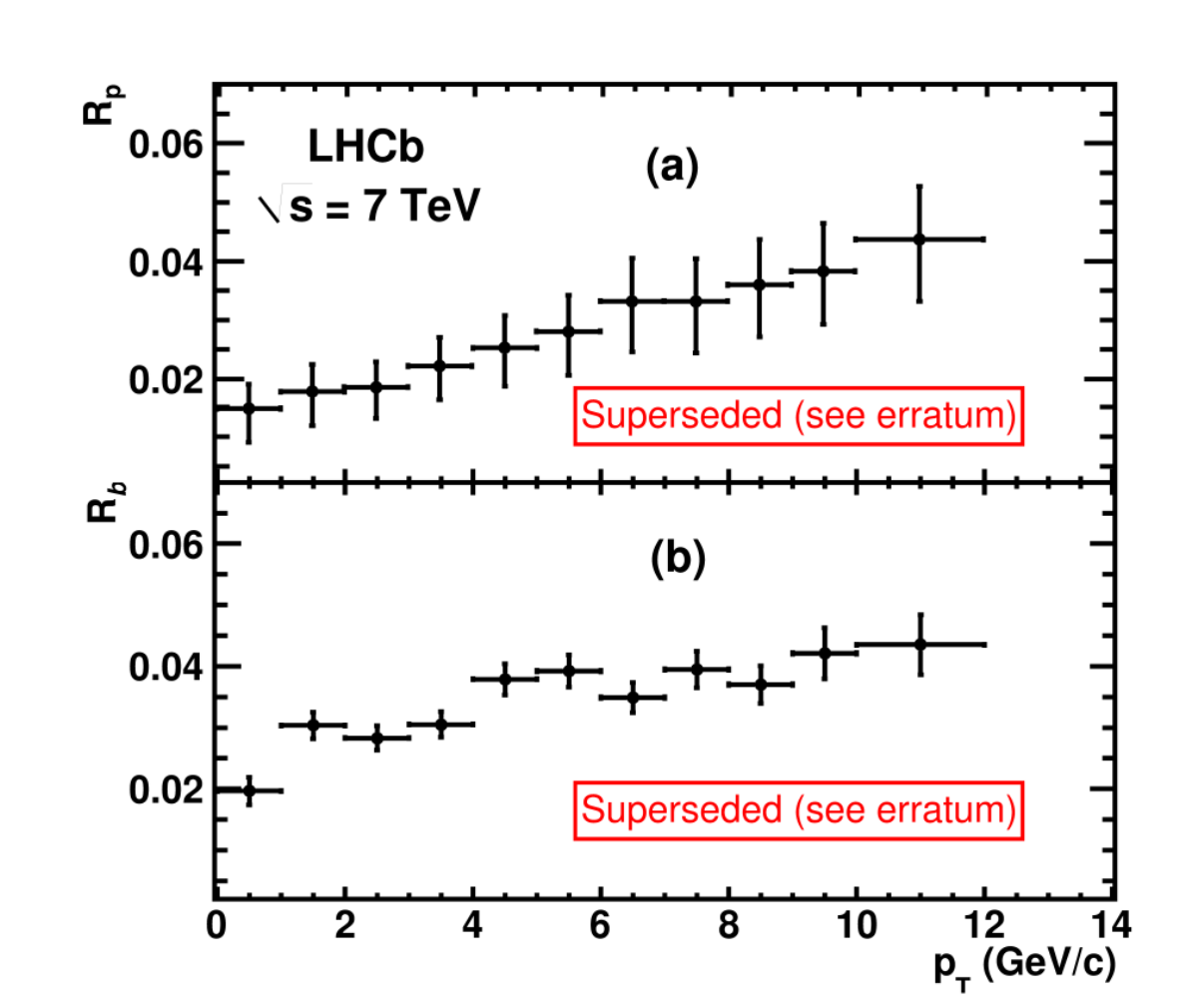}}
\caption{Ratio of $\psi(2S) \to \mu^+ \mu^-$ to $\jpsi \to \mu^+ \mu^-$ cross-sections for prompt 
production (a) and for $b$-hadron decay (b), as a function of \pt. \textbf{This figure is updated in the Erratum}.} 
\label{fig:ratio}
\end{center}
\end{figure}

The ratio of the $\tomm$ to $\toJmm$ differential cross-sections
is shown~{\em vs.} \pt in Fig.~\ref{fig:ratio} for 
prompt production ($R_{\rm p}$, Fig.~\ref{fig:ratio}(a)) and when the vector mesons originate 
from $b$-hadron decays ($R_{b}$, Fig.~\ref{fig:ratio}(b)). 
Since it is not known if the promptly produced \psitwos and \jpsi have
similar polarizations~\cite{bib:cdf2}, we do not assume any correlation of
the polarization uncertainties when computing the uncertainties on $R_{\rm p}$.
The increase of $R_{{\rm p}(b)}$ with \pt is similar to 
that measured in the central rapidity region by the CDF~\cite{bib:cdf}
and CMS~\cite{bib:cms} collaborations. \textbf{These results are updated by Sec. \ref{erratum:results}}.

\section{Conclusions}
We have measured the  differential cross-section 
for the process  $pp \to \psi(2S)X$ at the centre-of-mass energy 
of 7\tev, as a function of the transverse momentum in the range $\pt(\psi(2S))\le16$\gevc
and $2<y(\psi(2S))\le4.5$, via the decay modes \tomm and \topp.  
The data sample corresponds to about 36\invpb 
collected by the LHCb experiment at the LHC. 
Results from the two decay modes agree. 
The $\psi(2S)$ prompt cross-section has been separated 
from the cross-section of $\psi(2S)$ from $b$-hadrons through the
study of the pseudo-decay-time and the two measurements have been averaged. 
In the above kinematic range we measure
\begin{eqnarray*}
\sigma_{\rm prompt}(\psi(2S)) &=& 1.44 \pm 0.01~(\text{stat}) \pm 0.12~(\text{syst})^{+0.20}_{-0.40}~(\text{pol})~{\rm \upmu b}, \\
\sigma_{b}(\psi(2S)) &=& 0.25 \pm 0.01~(\text{stat}) \pm 0.02~(\text{syst})~{\rm \upmu b}. 
\end{eqnarray*}
The measured $\psi(2S)$ production cross-sections are in good agreement with the results
of several recent NRQCD calculations.
In addition, we obtain an improved value for the 
$b \rightarrow \psi(2S) X$ branching fraction
by combining the two LHCb production cross-section measurements of the two vector mesons 
$\jpsi$ and $\psi(2S)$ from $b$-hadrons. The result,
\begin{equation*}
\mathcal{B}(b \rightarrow \psi(2S) X) = (2.73 \pm 0.06~(\text{stat}) \pm 0.16~(\text{syst}) \pm 0.24~(\text{BF})) 
\times 10^{-3},
\end{equation*}
is in good agreement with recent results from the CMS collaboration~\cite{bib:cms} and is
a significant improvement over the present PDG average~\cite{bib:PDG}.

{\bf The results above are corrected by the Erratum in Appendix~\ref{erratum}.}

\section*{Acknowledgments}
We express our gratitude to our colleagues in the CERN accelerator
departments for the excellent performance of the LHC. We thank the
technical and administrative staff at CERN and at the LHCb institutes,
and acknowledge support from the National Agencies: CAPES, CNPq,
FAPERJ and FINEP (Brazil); CERN; NSFC (China); CNRS/IN2P3 (France);
BMBF, DFG, HGF and MPG (Germany); SFI (Ireland); INFN (Italy); FOM and
NWO (The Netherlands); SCSR (Poland); ANCS (Romania); MinES of Russia and
Rosatom (Russia); MICINN, XUNGAL and GENCAT (Spain); SNSF and SER
(Switzerland); NAS Ukraine (Ukraine); STFC (United Kingdom); NSF
(USA). We also acknowledge the support received from the ERC under FP7
and the Region Auvergne.

We thank B. Kniehl, M. Butensch\"{o}n and M. Cacciari for providing 
theoretical predictions of \psitwos cross-sections in the LHCb acceptance range.


\appendix
\def\xx {\ensuremath{\kern 0.5em }}
\def\mygevc {\ensuremath{{\mathrm{Ge\kern -0.1em V\!/}c}}\xspace}

\section{Erratum}\label{erratum}
This erratum corrects measurements of the prompt and secondary (from-$b$)
\psitwos~production cross-sections in the forward region
in $pp$ collisions at  \mbox{$\sqs=7\tev$}. The original measurements, reported in the body of this preprint, were performed using data collected with the LHCb detector in 2010 and were published in Ref.~\cite{LHCb-PAPER-2011-045}.
Corrected results for prompt \psitwos and \psitwos-from-\bquark in the kinematic range $\pt(\psitwos) < 16\gevc$ and $2.0 < y(\psitwos) < 4.5$ are 
\begin{equation*}
\begin{split}
\sigma_{\rm prompt}(\psitwos) &= 1.37 \pm 0.01~(\text{stat}) \pm 0.06~(\text{syst})^{+0.19}_{-0.38}~(\text{pol})~{\rm \upmu b}, \\
\sigma_{b}(\psitwos) &= 0.31 \pm 0.01~(\text{stat}) \pm 0.02~(\text{syst})~{\rm \upmu b}. 
\end{split}
\end{equation*}
where the last uncertainty on the prompt cross-section is due to 
the unknown $\psitwos$ polarization. 
With the corrected \psitwos-from-\bquark cross-section the inclusive branching fraction is updated by 
\begin{equation*}
\mathcal{B}(b \rightarrow \psitwos X) = (3.08 \pm 0.07(\mathrm{stat}) \pm 0.36(\mathrm{syst}) \pm 0.27(\mathrm{\BF})) \times 10^{-3}.
 \end{equation*}

\subsection{Nature of the correction}
\label{erratum:intro}

\setcounter{figure}{2}
\setcounter{table}{1}
\setcounter{equation}{0}

In Ref.~\cite{LHCb-PAPER-2011-045}, the production rate of \psitwos\ mesons
in the rapidity range $2.0 < y < 4.5$ was measured for
$pp$ collisions at \mbox{$\sqs=7\tev$}
using a sample of data corresponding to $36\invpb$.
Both overall and singly differential $(\mathrm{d}\sigma / \mathrm{d}\pt)$
cross-sections were measured by fitting the invariant-mass spectra to
obtain background-subtracted signal yields, which are subsequently efficiency corrected. Two decay modes were used:
\mbox{$\psitwos \to \mu^+ \mu^-$} and
\mbox{$\psitwos \to \jpsi (\mumu) \pip \pim$}.

Two sources of \psitwos\ production are expected in this environment:
mesons produced promptly in the primary interaction
(whether directly or through the decay of an intermediate resonance),
and those produced via the decays of $b$~hadrons.
The vast majority of $b$~hadrons produced in the LHCb acceptance
consist of $B^0$, $B^+$, $B_s^0$ mesons and \Lb baryons,
all with mean lifetimes of approximately $1.5$\,ps.
Consequently, the two classes of production may be separated according
to whether the \psitwos\ originates from the primary vertex (PV) or
from a downstream secondary vertex. This separation must be done on a
statistical level, since some $b$~hadrons will decay close to the PV
on the scale of the experimental resolution.

The pseudo-decay-time $t_z$ was used to distinguish the two sources of production,
and is defined as
\begin{equation}
t_z=\frac{\left(z_{\psitwos}-z_\mathrm{PV}\right)\times M_{\psitwos}}{p_z},
\end{equation}
where $z_{\psitwos}$ and $z_{\mathrm{PV}}$ are the $z$~coordinates
of the reconstructed $\psitwos$ decay vertex and the primary vertex,
$p_z$ is the $z$-component of the measured $\psitwos$ momentum, 
$M_{\psitwos}$ is the known $\psitwos$ mass~\cite{PDG2012},
and the $z$-axis is the direction of the proton beam pointing
downstream into the LHCb acceptance. For a given sample of $\psitwos$
candidates, a fit to the $t_z$ distribution was used to obtain
the prompt fraction $f_\mathrm{p}$, as described in Sec.~4 of Ref.~\cite{LHCb-PAPER-2011-045}.

Two distinct problems related to the determination of $f_\mathrm{p}$
in~Ref.~\cite{LHCb-PAPER-2011-045} have been identified.
The first is that a mathematical mistake was made in
calculating the systematic uncertainties
on the from-$b$ \psitwos\ production cross-sections
that arise due to uncertainties in the $t_z$ fit;
a factor of $f_\mathrm{p} / (1 - f_\mathrm{p})$ was omitted. When this mistake is
corrected, those systematic uncertainties increase by a factor 3 to 9, depending on the \pt defined in the range $0-16\gevc$,
with the largest effect at low \pt, where the prompt
fraction is close to unity. The correct formula is used in the
results below.

\begin{figure}[tb]
    \begin{center}
	\includegraphics[width=0.70\textwidth]{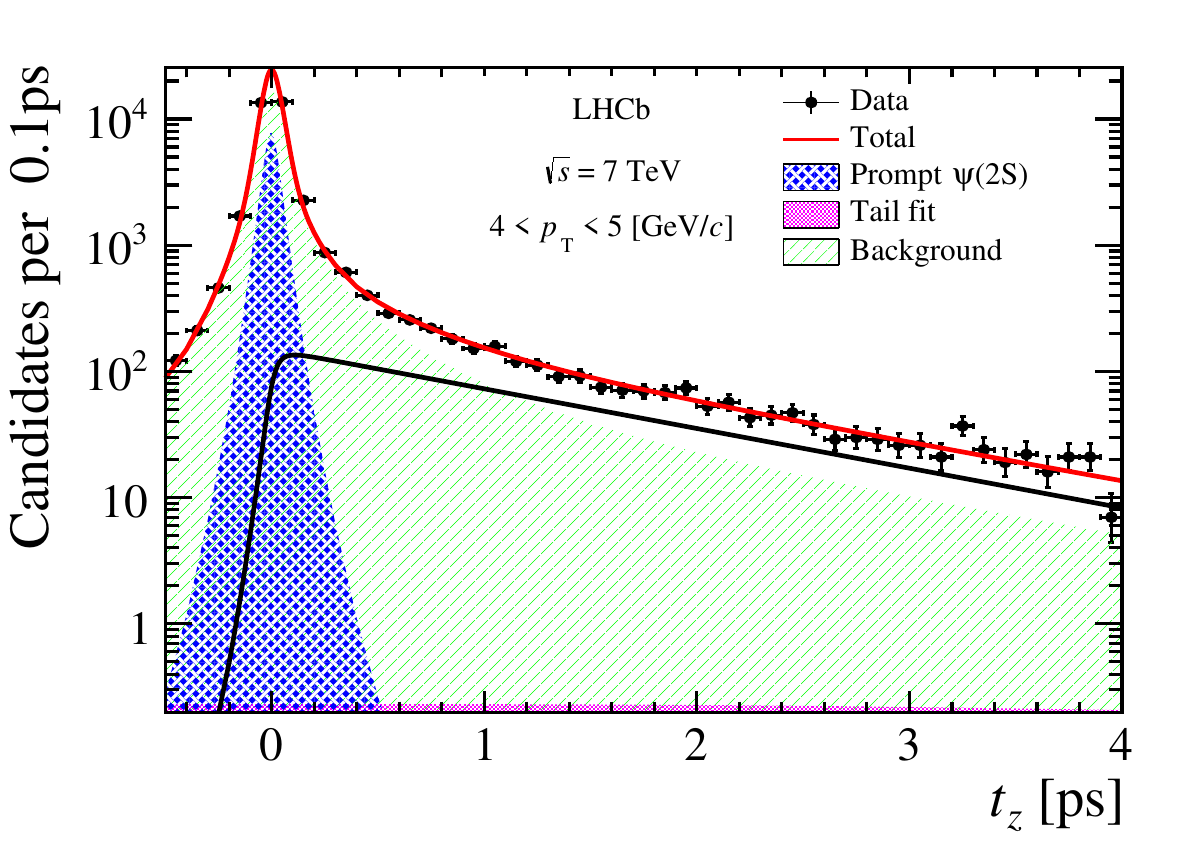}
	\caption{Pseudo-decay-time $t_z$ distribution for the \mbox{$\psitwos \to \mu^+ \mu^-$} decay mode in the range $4<\pt\le5$ \gevc, showing the background and prompt contributions.} 
	\label{fig:tz_fit_corrected}
	\end{center}
\end{figure}

The second problem is related to the values of $f_\mathrm{p}$ themselves.
A mistake appears to have been made in the measurement of $f_\mathrm{p}$
via the fits to the $t_z$ distributions used in Ref.~\cite{LHCb-PAPER-2011-045}.
An independent reimplementation of the analysis finds consistently lower values of $f_\mathrm{p}$. 
This change in $f_\mathrm{p}$ is associated with a change in the mean value of $t_z$ seen
for the from-$b$ component: values of approximately $1.1\ps$ were found in the analysis
reported in Ref.~\cite{LHCb-PAPER-2011-045}, compared to approximately $1.5\ps$
(much closer to the mean lifetime of contributing $b$~hadrons)
in the reimplementation. This issue has been found using both the original $t_z$ fit function as described in Ref.~\cite{LHCb-PAPER-2011-045} and the function used in Ref.~\cite{LHCb-PAPER-2018-049}. The more sophisticated $t_z$ fit function used in Ref.~\cite{LHCb-PAPER-2018-049} achieves a more precise description of experimental data and is thus used in obtaining corrections as described below. An example of a $t_z$ fit in the \pt range $4<\pt\le5$ \gevc is shown in Fig. \ref{fig:tz_fit_corrected}.

Because the issue found is limited to the determination of $f_\mathrm{p}$
and does not affect the combined cross-section,
and given that the reimplementation uses a sample of \mbox{$\psitwos \to \mumu$} events that
is correlated with but not identical to the original analysis,
the approach used in this erratum is to  use the new and old values of $f_\mathrm{p}$ to determine a correction factor to apply to the results of the prompt and from-$b$ cross-sections of the original analysis.
(A separate and statistically independent analysis of the
larger $7\tev$ data sample taken in 2011 has been submitted~\cite{LHCb-PAPER-2018-049}
but is outside the scope of this erratum.)
Defining $f_b \equiv 1 - f_\mathrm{p}$ for convenience, 
the ratio
\begin{equation}
  \mathcal{R}_{b} = \frac{ f_b\mbox{ obtained with reimplementation} }
             { f_b(\psitwos\to\mumu)\mbox{ obtained in original analysis} }
\end{equation}
is determined in bins of \pt. 
The correction is then obtained by fitting a linear function
to the individual values of $R_b$. This also allows the correction to be extrapolated to kinematic regions where data were not available for the reimplementation ($\pt < 2\gevc$, $\pt > 11\gevc$). 
This correction is applied to the weighted average of \mbox{$\psitwos \to \mumu$} and \mbox{$\psitwos \to \jpsi (\mumu) \pip \pim$} results as reported by Ref.~\cite{LHCb-PAPER-2011-045}.

After applying the correction to $f_b$,
the systematic uncertainties are recomputed.
These are unchanged respect to those of the original analysis (other than
relative uncertainties being updated for the new central values)
except as described below. First, the the mistake in the computation of the uncertainty associated with the $t_z$ fit is corrected as described above.
Second, a new systematic uncertainty associated with the $f_b$
correction estimate is added, and in particular the extrapolation outside
the fit region, which is determined by taking the difference
between the correction fitted by a first-order and a second-order
polynomial.

\subsection{Corrected results}
\label{erratum:results}

The impact on $f_\mathrm{p}$ itself and on the cross-section
for prompt production is modest: they are both reduced by an amount typically of the order
of several percent. However the relative impact on $f_b$ is greater, and the from-$b$ cross-section
rises by typically 20--25\%.

\begin{figure}
    \begin{center}
	\includegraphics[width=0.70\textwidth]{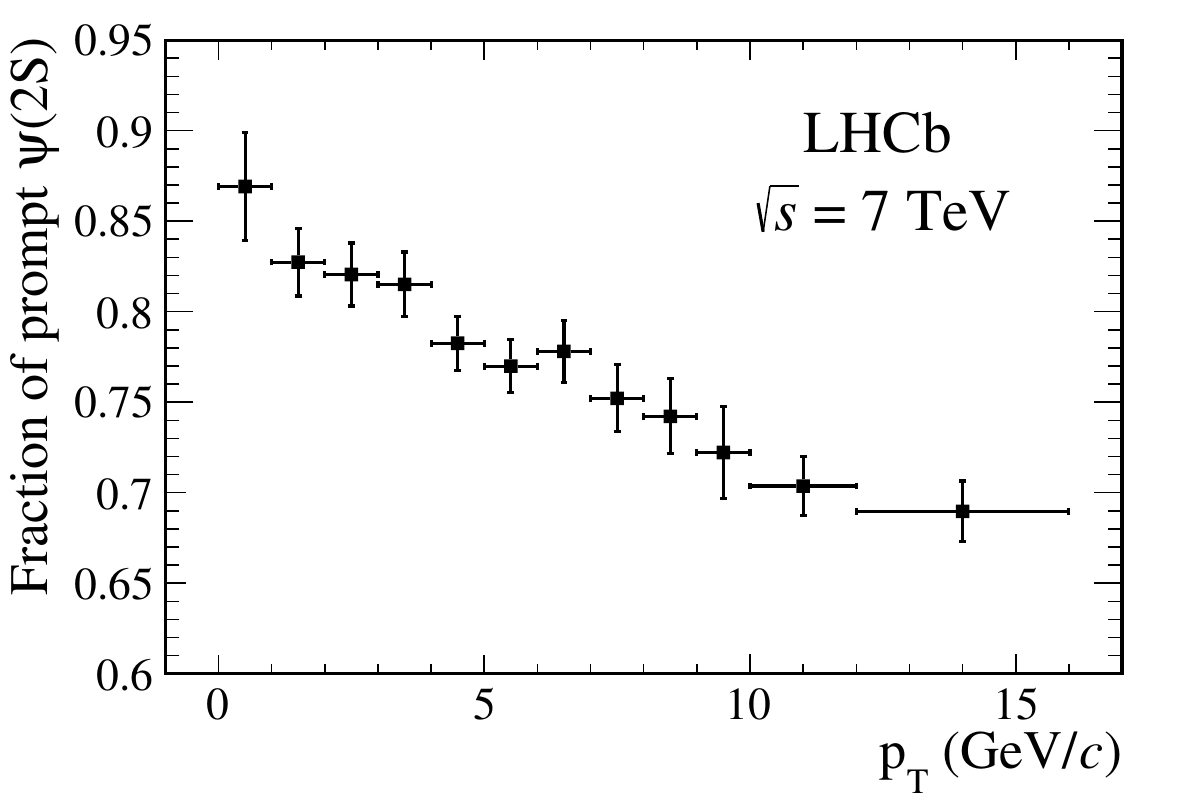}
	\caption{Fraction of prompt \psitwos, $f_\mathrm{p}$, as a function of \pt. The error bars include statistical and systematic uncertainties added in quadrature.} 
	\label{fig:Fp_corrected}
	\end{center}
\end{figure}

Corrected versions of all figures and tables 
in Ref.~\cite{LHCb-PAPER-2011-045}
that were affected by the issue are given in the following.
The corrected $f_\mathrm{p}$ distribution as a function of \pt~is shown in Fig.~\ref{fig:Fp_corrected}.
The singly differential cross-section as a function of \pt\,is shown for prompt production in Fig.~\ref{fig:results_prompt},
and for production from $b$-hadrons in Fig.~\ref{fig:results_fromb}.
In the figures, the updated cross-sections are compared with theory predictions,
namely NRQCD calculations~\cite{Shao:2014yta} for prompt production
and FONLL calculations\cite{Cacciari:1998it} for production of \psitwos\ from
$b$-hadron decays. The integrated cross-sections in the nominal kinematic range for prompt \psitwos and \psitwos-from-\bquark are found to be
\begin{equation*}
\begin{split}
\sigma_{\rm prompt}(\psitwos) &= 1.37 \pm 0.01~(\text{stat}) \pm 0.06~(\text{syst})^{+0.19}_{-0.38}~(\text{pol})\mub, \\
\sigma_{b}(\psitwos) &= 0.31 \pm 0.01~(\text{stat}) \pm 0.02~(\text{syst})\mub. 
\end{split}
\end{equation*}
The numerical results are given in Table~\ref{tab_erratum:results_PT}.

Corrected ratio of \mbox{$\psitwos \to \mu^+ \mu^-$} and \mbox{$\jpsi \to \mumu$} cross-sections for prompt production ($\mathrm{R_{p}}$) and for \textit{b}-hadron decay ($\mathrm{R}_{b}$) as a function of \pt is shown on Fig. \ref{fig:ratio_psi2S_jpsi}.

The inclusive $\bquark\to\psitwos X$ branching fraction is computed using the \psitwos-from-\bquark cross-sections reported above
and found to be

\begin{equation*}
\mathcal{B}(b \to \psitwos X) = (3.08 \pm 0.07(\mathrm{stat}) \pm 0.36(\mathrm{syst}) \pm 0.27(\mathrm{\BF})) \times 10^{-3}.
 \end{equation*}
The last uncertainty is due to those of the branching fractions, and is dominated by the $\BF(b \to \jpsi X)$ uncertainty.

\setcounter{figure}{6}

\begin{figure}
    \begin{center}
	\includegraphics[width=0.70\textwidth]{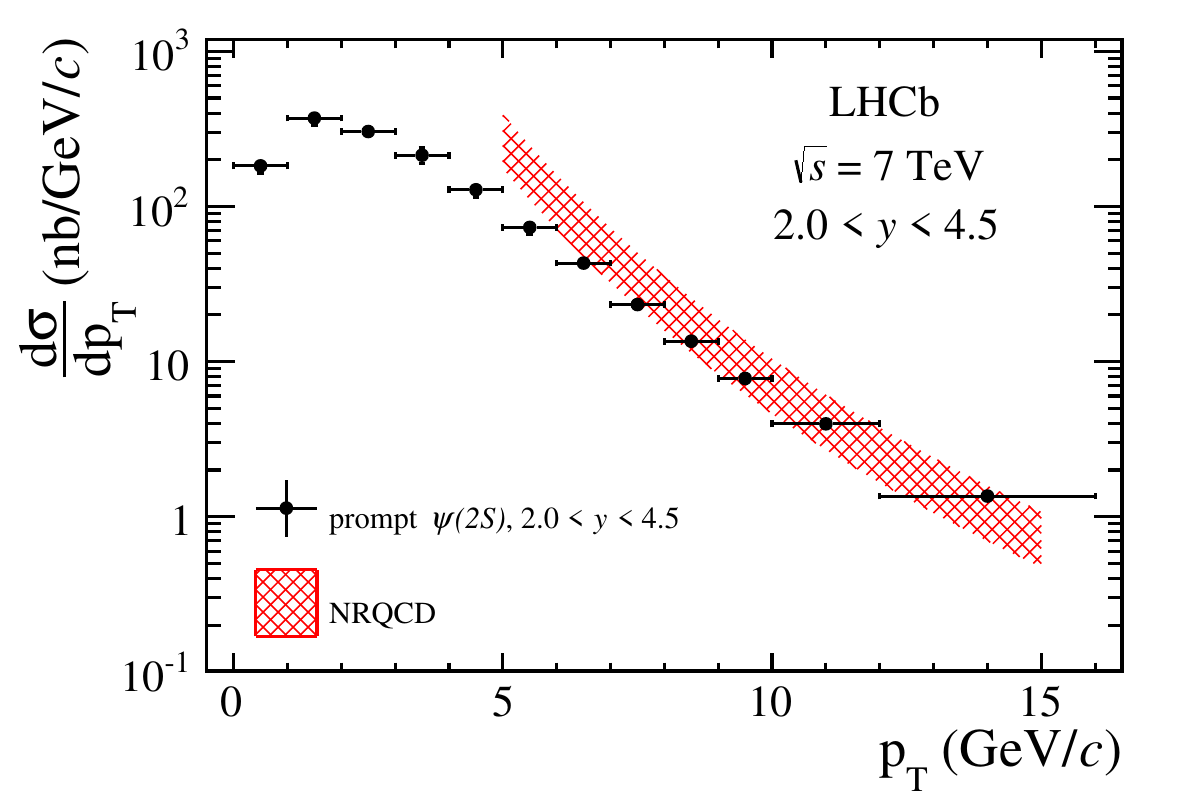}
	\caption{Differential production cross-section of prompt \psitwos~as a function of \pt in the range $2.0 < y < 4.5$. The results are compared with the NRQCD calculations~\cite{Shao:2014yta}.  The error bars include statistical and systematic uncertainties added in quadrature.} 
	\label{fig:results_prompt}
	\end{center}
\end{figure}

\begin{figure}
    \begin{center}
	\includegraphics[width=0.70\textwidth]{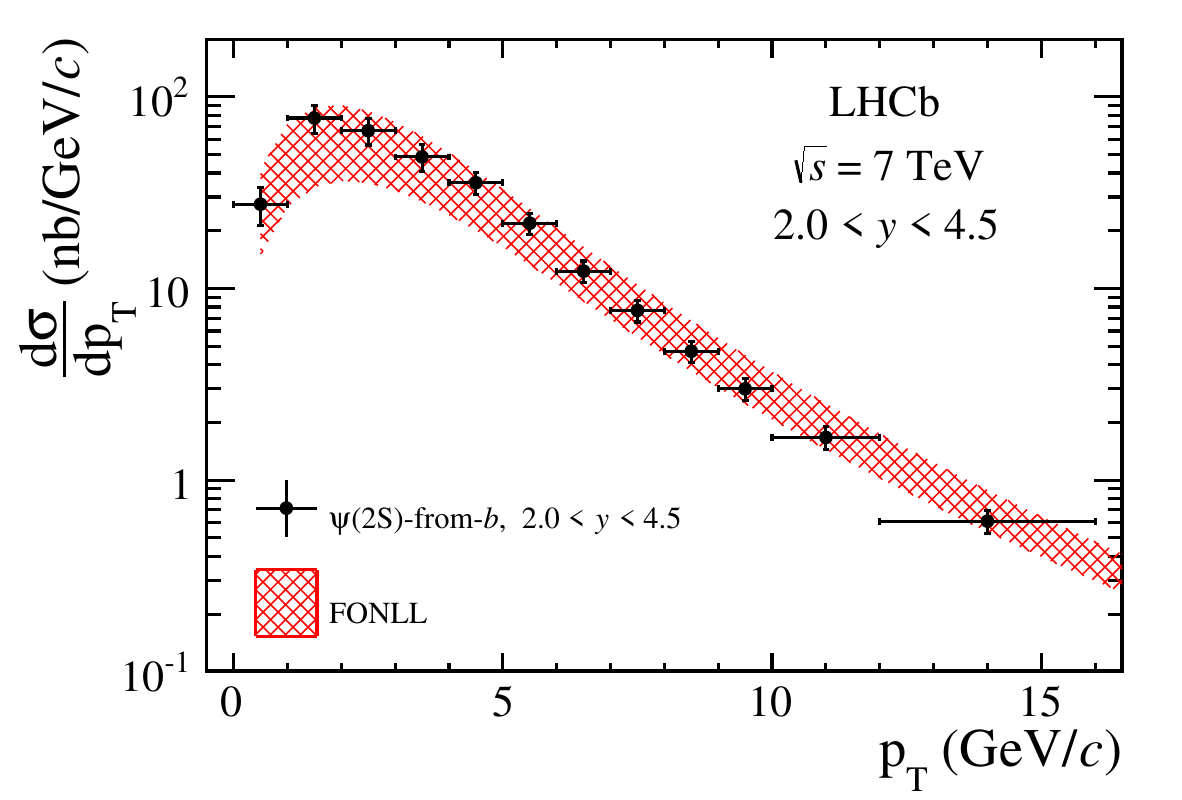}
	\caption{Differential production cross-section of \psitwos from $b$~hadrons as a function of \pt in the range $2.0 < y < 4.5$. The results are compared with the FONLL calculations~\cite{Cacciari:1998it}. The error bars include statistical and systematic uncertainties added in quadrature.} 
	\label{fig:results_fromb}
	\end{center}
\end{figure}

\begin{figure}[!tbp]
\centering
\begin{minipage}[t]{0.49\textwidth}
\centering
\includegraphics[width=1.0\textwidth]{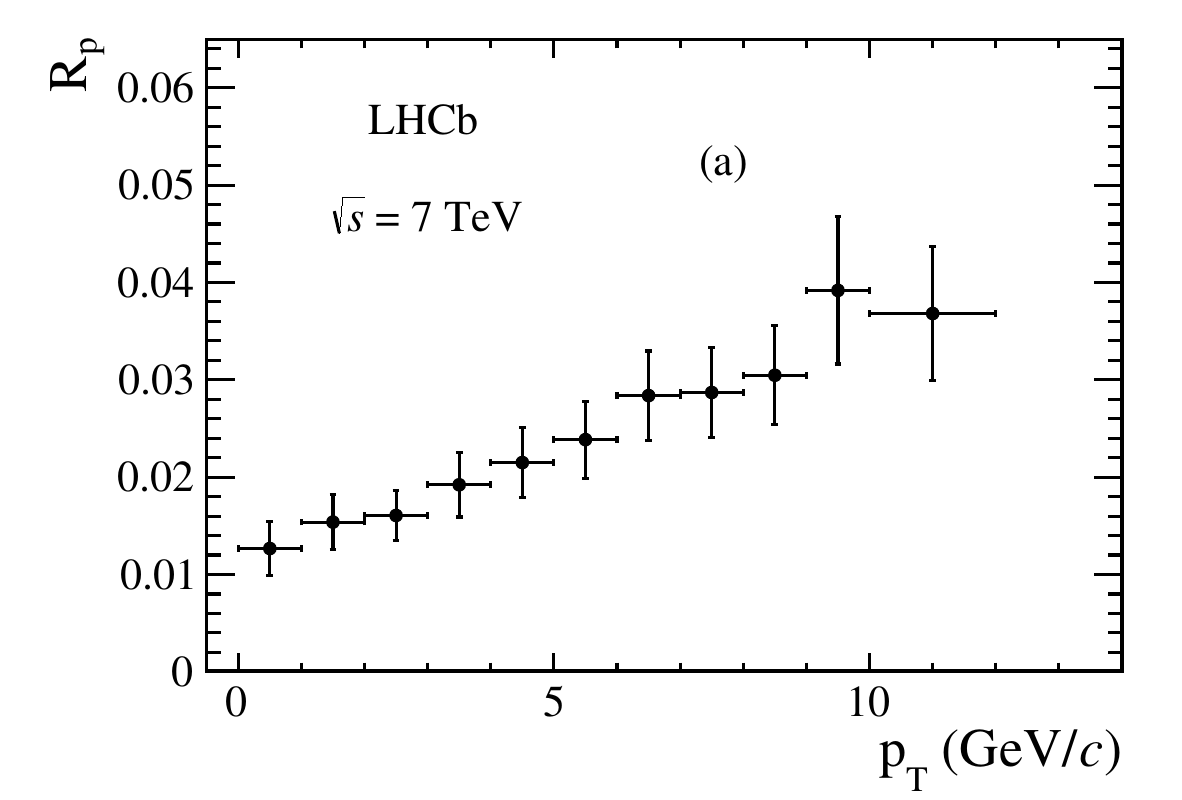}
\end{minipage}
\begin{minipage}[t]{0.49\textwidth}
\centering
\includegraphics[width=1.0\textwidth]{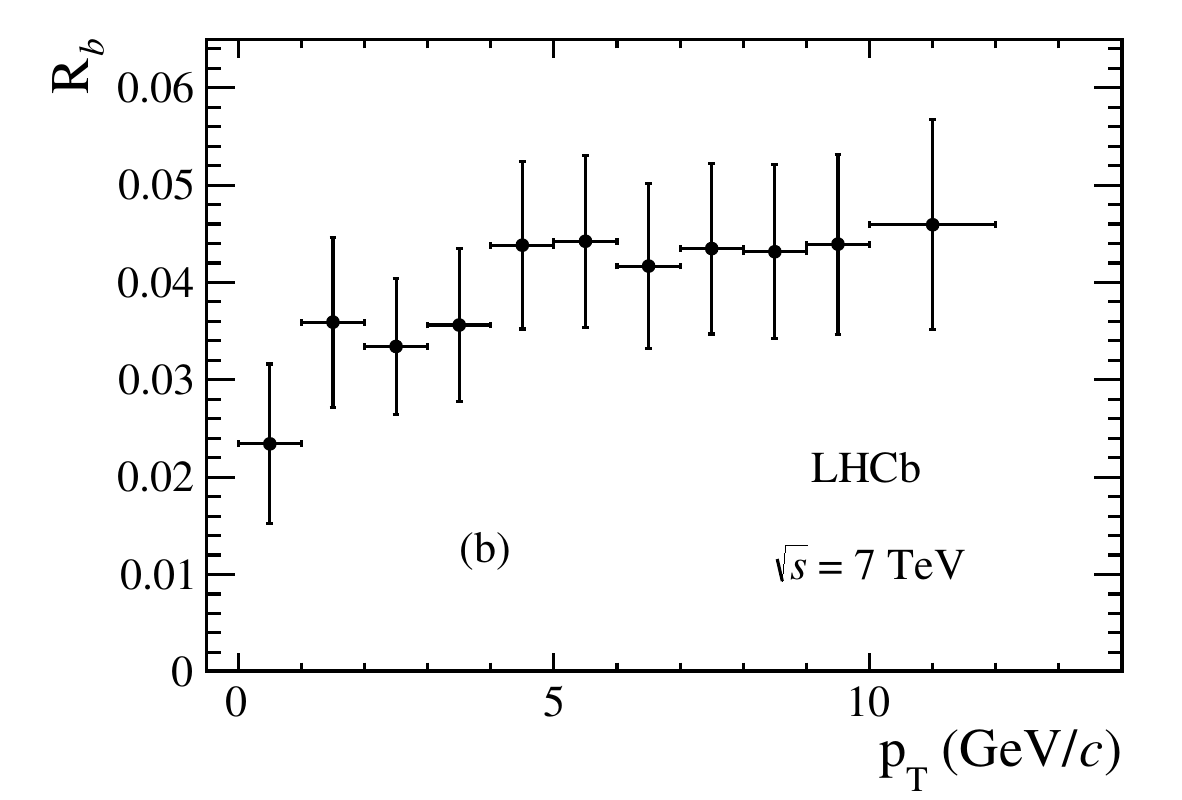}
\end{minipage}
\caption{
    Ratio of \mbox{$\psitwos \to \mu^+ \mu^-$} and \mbox{$\jpsi \to \mumu$} cross-sections for prompt production (a) and for \textit{b}-hadron decay (b), as a fucntion of \pt.}
\label{fig:ratio_psi2S_jpsi}
\end{figure}

\begin{table}[htp]
\caption{Differential cross-sections $\deriv\sigma/\deriv\pt$ (in \nb/(\mygevc)) of prompt \psitwos and \psitwos-from-$b$ hadrons at $\sqrt{s} = 7\tev$, integrated over $y$ between 2.0 and 4.5. The first uncertainty is statistical and the second systematic. The third asymmetric uncertainty for the prompt \psitwos mesons is due to the unknown polarisation.}
\centering
\setlength{\tabcolsep}{10pt} 
\renewcommand{\arraystretch}{1.5} 
\begin{tabular}{c|cc}
\hline
\pt(\mygevc)& Prompt \psitwos & \psitwos-from-$b$ \\
\hline
\kern 0.75em0--1 & $183\pm6\pm18~^{+31}_{-65}$ & $28\pm3\pm6$\\
\kern 0.75em1--2 & $\xx371\pm7\pm37~^{+58}_{-114}$ & $\xx77\pm4\pm13$\\
\kern 0.75em2--3 & $304\pm6\pm26~^{+42}_{-84}$ & \xx$67\pm3\pm10$\\
\kern 0.75em3--4 & $214\pm6\pm24~^{+26}_{-51}$ & $49\pm3\pm8$\\
\kern 0.75em4--5 & $128\pm4\pm13~^{+15}_{-29}$ & $36\pm2\pm4$\\
\kern 0.75em5--6 & $73\pm2\pm7~^{+9}_{-17}$ & $22\pm1\pm3$\\
\kern 0.75em6--7 & $43\pm1\pm4~^{+5}_{-9}$ & $12\pm1\pm1$\\
\kern 0.75em7--8 & $23\pm1\pm2~^{+3}_{-6}$ & $\xx7.7\pm0.5\pm0.9$\\
\kern 0.75em8--9 & $14\pm1\pm1~^{+2}_{-3}$ & $\xx4.7\pm0.3\pm0.5$\\
\kern 0.75em9--10 & $\xx\xx7.8\pm0.4\pm0.7~^{+0.8}_{-1.6}$ & $\xx3.0\pm0.3\pm0.3$\\
\kern 0.25em10--12 &$\xx\xx4.0\pm0.2\pm0.4~^{+0.5}_{-0.7}$ & $\xx1.7\pm0.2\pm0.2$\\
\kern 0.25em12--16 & $\xx\xx1.4\pm0.1\pm0.2~^{+0.2}_{-0.3}$ & $\xx0.61\pm0.05\pm0.07$\\
 \hline
\kern 0.75em0--16 & $\xx\xx1366\pm13\pm56~^{+190}_{-380}$ & $\xx308\pm6\pm19$
\end{tabular}
\label{tab_erratum:results_PT}
\end{table}

\clearpage

\addcontentsline{toc}{section}{References}
\bibliographystyle{LHCb}
\bibliography{main,standard,LHCb-PAPER,LHCb-CONF,LHCb-DP,LHCb-TDR,local}

\newpage
\centerline{\large\bf LHCb collaboration}
\begin{flushleft}
\small
R.~Aaij$^{23}$,
C.~Abell{\'a}n~Beteta$^{35}$,
B.~Adeva$^{36}$,
M.~Adinolfi$^{43}$,
C.~Adrover$^{6}$,
A.~Affolder$^{49}$,
Z.~Ajaltouni$^{5}$,
J.~Albrecht$^{37}$,
F.~Alessio$^{37}$,
M.~Alexander$^{48}$
% ...

{\footnotesize \it
% ...
$ ^{5}$Universit{\'e} Clermont Auvergne, CNRS/IN2P3, LPC, Clermont-Ferrand, France\\
$ ^{6}$Aix Marseille Univ, CNRS/IN2P3, CPPM, Marseille, France\\
% ...
$ ^{23}$Nikhef National Institute for Subatomic Physics, Amsterdam, Netherlands\\
% ...
$ ^{35}$ICCUB, Universitat de Barcelona, Barcelona, Spain\\
$ ^{36}$Instituto Galego de F{\'\i}sica de Altas Enerx{\'\i}as (IGFAE), Universidade de Santiago de Compostela, Santiago de Compostela, Spain\\
$ ^{37}$European Organization for Nuclear Research (CERN), Geneva, Switzerland\\
% ...
$ ^{43}$H.H. Wills Physics Laboratory, University of Bristol, Bristol, United Kingdom\\
% ...
$ ^{48}$School of Physics and Astronomy, University of Glasgow, Glasgow, United Kingdom\\
$ ^{49}$Oliver Lodge Laboratory, University of Liverpool, Liverpool, United Kingdom\\
% ...
}
\end{flushleft}
\end{document}